\documentclass[a4paper,superscriptaddress,prb, showkeys]{revtex4}
\usepackage[latin1]{inputenc}
\usepackage{bbm}
\usepackage{graphicx}
\usepackage{cancel}
\usepackage{lmodern}
\usepackage{color}
\usepackage{bbm}
\usepackage{lipsum}
\usepackage{amsmath}
\usepackage{amsfonts}
\usepackage{amssymb}
\usepackage{mathrsfs}
\newcommand\beq{\begin{equation}}
\newcommand\eeq{\end{equation}}
\newcommand\bear{\begin{eqnarray}}
\newcommand\eear{\end{eqnarray}}
\usepackage{epstopdf}
\usepackage{pdfpages}
\usepackage{setspace}
\usepackage{color}

\usepackage{tabularx,ragged2e,booktabs,caption}
\begin{document}
\title{\center Invariant-theoretic approach to nonlinear hyperelastic constitutive modeling of graphene}
\author{Sandeep Kumar }
\email{lahirisd@mit.edu}
\affiliation{ Department of Mechanical Engineering \\ Massachusetts Institute of Technology, Cambridge, MA 02139}
\author{David M. Parks}
\email{dmparks@mit.edu}
\affiliation{ Department of Mechanical Engineering \\ Massachusetts Institute of Technology, Cambridge, MA 02139}
\begin{abstract}
 We develop a hyperelastic constitutive model for graphene --- describing in-plane deformations involving both large isotropic and deviatoric strains --- based on the invariant-theoretic approach to representation of anisotropic functions. The strain energy density function $\psi$ is expressed in terms of special scalar-valued functions of the 2D logarithmic strain tensor $ \mathbf E^{(0)}$ --- called  the symmetry-invariants --- that remain invariant w.r.t. the material symmetry group of graphene, $\mathcal G = \mathcal C_{6v}$.  Our constitutive model conforms to a larger set of \textit{ab initio} energies/stresses while introducing fewer elastic constants than previously-proposed models. In particular, when the strain energy is expressed in terms of symmetry invariants, the material symmetry group is intrinsically incorporated within the constitutive response functions; consequently, the elastic constants (of all orders) in the formulation are \textit{a priori} independent: i.e., no two or more elastic constants are related by symmetry. This offers substantial simplification in terms of formulation as it eliminates the need for identifying the independent elastic constants, a task which can become particularly cumbersome as higher-order terms in the strain energy density function are incorporated. We validate our constitutive model by computing (1) the stress and (2) the elastic stability limit  for a set of homogeneous finite deformations --- comprising uniaxial  stretch/stress along the armchair, and the zigzag directions; and equi-biaxial tension. The stress values predicted by the model are in good agreement with the directly-calculated \textit{ab initio} values. The elastic stability limits predicted by acoustic tensor analysis compare well with the predictions from phonon calculations carried out independently using linear response density functional perturbation theory.
\end{abstract}
\maketitle
\section{Introduction}
 The anisotropy of an ideal crystalline material is characterized by its material symmetry group $\mathcal G$--- the set of all symmetry operations about a point that leave the crystal unchanged.  The material symmetry group is reflected in all physical properties of the crystalline material- microscopic as well as macroscopic. For example, in  a two-dimensional (2D) crystal like graphene, the strain energy density function $\psi$, which relates the change in free energy per unit area ($\psi$) to the measure of elastic strain ($\mathbf E^{(0)}$), --- possesses the complete point group symmetry  of the underlying lattice, i.e., $\hat{\psi}(\mathbf E^{(0)}) =\hat{\psi}(\mathbf Q^T \mathbf E^{(0)} \mathbf Q)\ \forall \mathbf Q \in \mathcal G$.
  Therefore, in any constitutive model of the crystal, all symmetries belonging to the material symmetry group should be duly incorporated \cite{Neumann}.\\
 One approach to embed the material symmetry into constitutive modeling is \textit{reduction by symmetry}. In this approach to hyperelastic constitutive modeling of anisotropic crystals, $\psi$ is expressed as a power series of desired order in components of the strain tensor, and then  symmetry constraints are imposed to reduce the number of independent fitting coefficients \cite{wallace1998thermodynamics}.\\
Cadelano \textit{et al.} \cite{cadelano2009nonlinear} employed this method  to develop a hyperelastic nonlinear constitutive model of monolayer graphene. The strain energy density function $\psi$ in this model is expressed as a  cubic polynomial in components of the Green-Lagrange strain measure, and hence exhibits a quadratic nonlinearity in the work-conjugate stress-strain relation.  Upon reduction by symmetry, the expression for  $\psi$ contains three independent third-order elastic constants, and two second-order elastic constants, the values of which were determined by a least-squares fit to the strain energies obtained from tight-binding atomistic simulations for a set of canonical deformations comprising uniaxial stretch along the armchair and the zigzag directions; an in-plane shear; and biaxial tension.\\
Wei \textit{et al.} \cite{wei2009nonlinear, lee2008measurement} noted that the model of Cadelano \textit{et al.} failed to reproduce the response at both infinitesimal and finite strains simultaneously with sufficient fidelity. They proposed  a fifth-order series expansion for 
$\psi$, again based on components of the Green-Lagrange strain measure,  to model the in-plane elastic properties of graphene. Their symmetry-reduced expression contained fourteen independent elastic constants, the values of which were determined by fitting model results to  DFT-calculated stresses corresponding to the same set of deformations used by Cadelano \textit{et al.}\\
The reduction by symmetry approach suffers from a number of disadvantages: First, the method requires imposing the symmetry constraints extrinsically on components of the elastic constant tensors. This procedure must be repeated each time a higher-order polynomial nonlinearity is added in the model. Since the number of components in a 2D tensor increases with its order $m$ as $2^m$, the task of reduction by symmetry becomes increasingly tedious as the order of the polynomial expansion increases. 
Secondly, the symmetry restrictions are most readily identified and applied to strain components evaluated in a particular crystallographic frame. From an implementation point of view, it would be preferable to express $\psi$  in a fashion independent of any particular choice of coordinate axes.
This can  best be accomplished  if $\psi$ is expressed as a function of a list of invariants of the strain measure, instead of its components.  Lastly, and most importantly, the method is \textit{a priori} restricted to the use of polynomials in the components of strain as the basis of representation --- with no scope for including arbitrary non-polynomial functions in the representation. \\
Here we present a constitutive modeling scheme for crystal hyperelasticity based on symmetry invariants of the logarithmic measure of strain $\mathbf E^{(0)}$. The \textit{symmetry invariants} are scalar-valued functions of the tensor agency (here, strain) that satisfy all the symmetry constraints of the crystal's material symmetry group \cite{Smith1,Smith2,Smith3}. 
Using the symmetry-invariants of $\mathbf E^{(0)}$ for $C_{6v}$ symmetry, we obtain a constitutive model of graphene for deformations involving both large area-change and shape-change. The resulting description of material response automatically satisfies the material symmetry requirements without further constraints.
 The model can be readily extended to include higher-order nonlinearities of either polynomial or functional natures.  The approach enables straightforward evaluation of the stress tensor, the tangent moduli tensor and the acoustic tensor in a coordinate-independent form.\\
\hspace*{0.5cm}The outline of this paper is as follows. In Sec.~\ref{kine}, we briefly discuss the kinematics of a 2D deformable body, and  introduce  relevant field variables and notations  used in the formulation.  In Sec.~\ref{inv}, we outline the general framework of hyperelasticity, Boehler's principle of isotropy of space based on the structure tensor \cite{boehler1977irreducible,boehler1979simple}, and  invariant-based  representation theory \cite{Smith1,Smith2,Smith3}. The structure tensor and  symmetry-invariants of $\mathbf E^{(0)}$ are explicitly derived for graphene in 
Sec.~\ref{hyper}. Form-invariant tensors are obtained as the derivatives of the symmetry-invariants with respect to $\mathbf E^{(0)}$.  
In Sec.~\ref{rep}, we propose a representation for $\psi$ in the basis of the symmetry-invariants. The expression for work-conjugate stress in terms of the  form-invariants  and the conversion from the work-conjugate stress measure to Cauchy stress is presented  in Sec.~\ref{wconj}. We obtain expressions for the work-conjugate tangent modulus tensor in section \ref{tang}, and for the acoustic tensor in Sec.~\ref{acou}. Section~\ref{train} describes the density functional theory (DFT) calculations done to provide a training dataset for evaluating material constants in the new constitutive model. The model is validated in Sec.~\ref{valid} by computing (1) the stress and (2) the elastic stability limits  for a set of finite deformations comprising uniaxial  stretch/stress along the armchair and the zigzag directions; and equi-biaxial tension. We conclude  in Sec.~\ref{summ} by summarizing our results and reflecting upon some implications.
\section{Kinematics}
\label{kine}
We consider graphene as a 2D deformable body denoted by unstressed reference configuration $\mathcal B$. The kinematics of this  deformable body is described by time-varying vector and tensor fields  belonging to the  2D Euclidian space $\mathbb R^2$.  We denote by $\mathbf X$ an arbitrary material point of $\mathcal B$. As the body is deformed, the material point $\mathbf X$ moves to another point in the 2D space, characterized by its deformed coordinate $\mathbf x$ at current time $t$. The convection of material points under deformation is described by a smooth, injective (one-to-one) function $\chi(\mathbf X, t)$ called the motion. The non-translational part of the motion can be equivalently defined by the positive-definite  second-order deformation gradient tensor, $\mathbf F = \nabla \chi(\mathbf X, t) $. 
Notationally suppressing this functional dependence for convenience, the polar decomposition theorem provides the following factorizations of $\mathbf F$\cite{gurtin2010mechanics, gurtin1982introduction, fung1977first}:
\begin{equation}
\mathbf F= \mathbf R \mathbf U =   \mathbf V\mathbf R,
\end{equation}
where the orthogonal tensor $\mathbf R \in SO_2$ characterizes  rigid-body rotation, 
whereas $\mathbf U$ (or $\mathbf V= \mathbf{RUR}^{T}$), termed the right (left) Cauchy-Green tensor, characterizes shape- and area-change. Physically,  deformation in the neighborhood of a material point in the body can be kinematically considered as stretching followed by a superimposed rigid-body rotation, or vice-versa. $\mathbf U$ is  a symmetric tensor having two real positive eigenvalues, the principal stretches $\lambda_1$ and $\lambda_2$.  Using  spectral decomposition, $\mathbf U$ can be expressed as
\begin{equation}
\mathbf U = \lambda_1 \mathbf r_1 \otimes \mathbf r_1 +\lambda_2 \mathbf r_2 \otimes \mathbf r_2,
\end{equation} 
where $\mathbf r_1$ and $\mathbf r_2$ are orthogonal principal unit vectors in the plane. The deformation of a material point can be kinematically factored  as the product of a purely dilatational (or  shape-preserving, but area-changing) deformation $\mathbf U^{a}$, and a purely isochoric (or 
shape-changing, but area-preserving) deformation $\tilde{\mathbf U}$. Accordingly, the stretch tensor can be product-decomposed as
\begin{equation}
      \mathbf U = \mathbf{U}^{a} \tilde{\mathbf U} =  \tilde{\mathbf U} \mathbf{U}^{a},
\end{equation}
where
\begin{equation}
   \mathbf{U}^{a}  \equiv J^{1/2} \, \mathbf{I}
\end{equation}
and 
\begin{equation}
   \tilde{\mathbf U} \equiv \lambda \mathbf r_1 \otimes \mathbf r_1 +\lambda^{-1}\mathbf r_2 \otimes \mathbf r_2;
\end{equation}
here $J = \mathrm{det}\, \mathbf U = \lambda_1 \lambda_2,\ \lambda = \sqrt{\lambda_1/\lambda_2}\ge 1$, and $\mathbf I$ is the 2D identity tensor.

The spectral representation of  $\mathbf E^{(0)} \equiv \ln \mathbf U = \ln \mathbf U^{a} + \ln \tilde{\mathbf U}$  is then given by:
\begin{eqnarray}
\mathbf E^{(0)} &=& \underbrace{\frac{1}{2}\,  \ln J \, \mathbf I}_{  \ln  \mathbf U^{a}} + \underbrace{\ln \lambda \, \left(\mathbf r_1 \otimes \mathbf r_1 - \mathbf r_2 \otimes \mathbf r_2 \right)}_{  \ln \, \tilde{\mathbf U}}  \nonumber \\ 
                    & \equiv &  \overbrace{\, \frac{1}{2} \, \, \epsilon_a \, \mathbf{I}}  \quad +   \overbrace{\quad \quad  \quad \quad \mathbf{E}^{(0)}_{0} ,\quad \quad \quad}
 \label{log}
\end{eqnarray}
where  
\begin{equation}
\epsilon_a = \text{tr}\, \mathbf E^{(0)} = \ln J = \ln (\det\, \mathbf U),
\end{equation}
gives the areal logarithmic strain $\epsilon_a$, and 
\begin{equation}
\mathbf E^{(0)}_0 = \ln \tilde{\mathbf U} = \ln \lambda \, \left(\mathbf r_1 \otimes \mathbf r_1 - \mathbf r_2 \otimes \mathbf r_2 \right),
\label{sprep}
\end{equation}
denotes the deviatoric part of  $\mathbf {E}^{(0)}$.

\section{Invariant-theoretic approach to constitutive modeling}
\label{inv}
In a hyperelastic modeling framework, the strain energy density $\psi$ is formally expressed as a scalar-valued function of deformation gradient $\mathbf F$. Material frame indifference requires that this function should remain invariant under superimposed rigid body motion; i.e., 
\begin{equation}
\psi = \hat \psi (\mathbf F) =  \hat \psi (\mathbf Q \mathbf F),
\end{equation} where arbitrary $\mathbf Q \in SO_2$ denotes a rigid-body rotation. 
Such objectivity is automatically satisfied if $\psi$ is functionally dependent on one of the Seth-Hill strain measures --- $\mathbf E^{(m)} \equiv ((\mathbf F^T \mathbf F)^{m/2} - \mathbf I)/m$, for $m \in \mathbb Z$ ($\mathbb Z$ is the set of real numbers). In particular, for $m=0$, the corresponding strain measure is the logarithmic strain measure $\mathbf E^{(0)}= \ln \mathbf U$ \cite{hencky1933elastic, hencky1931law}.\\
The choice of a particular strain measure is, in principle, totally arbitrary.  However,  different choices of strain measure can lead to differing levels of complexity or simplification in accurately  describing  material response. For example,  Anand \cite{anand1979h} 
explored the extension of the classical quadratic strain energy function of isotropic linear elasticity based on two small-strain Lam\'{e} constants by systematically replacing the invariants of the infinitesimal strain tensor with the corresponding invariants of
various of the Seth-Hill strain measures. For moderately large deformations, the formulation based on invariants of $\mathbf E^{(0)}$ (Hencky's strain energy function) most accurately captured the initial constitutive non-linearities, with results clearly superior to those obtained by similarly using the invariants of the Green-Lagrange strain measure $\mathbf E^{(2)}$.\\ 
Here, for reasons further described in Sec. \ref{rep}, we employ the logarithmic strain tensor $\mathbf E^{(0)}$ to write a hyperelastic constitutive response of graphene, i.e., 
\begin{equation}
\psi = \hat \psi(\mathbf F) = \bar \psi(\mathbf E^{(0)}).
\end{equation}
The restrictions on $\psi$ due to material symmetry are expressed as  
\begin{equation}
\bar \psi (\mathbf E^{(0)}) = \bar \psi(\mathbf Q^T \mathbf E^{(0)} \mathbf Q)\  \forall \  \mathbf Q \in \mathcal G, 
\label{isotropicization}
\end{equation} 
where $\mathcal G$ denotes the material symmetry group. A 
 scalar function, such as $\psi$,  that remains invariant under a  material symmetry group $\mathcal G$ is called a $\mathcal G$-invariant scalar function. Obtaining a representation for a generic $\mathcal G$-invariant function involves using the isotropicization theorem and symmetry invariants.\\ 
 The isotropicization theorem --- based on the notion of a materially-embedded structure tensor $\mathbb H$ --- allows the constitutive response of an anisotropic hyperelastic material to be expressed in terms of a list of special functions --- $\mathcal J_1$, $\mathcal J_2$, ..., $\mathcal J_n$ --- which are joint isotropic functions of $\mathbf E^{(0)}$ and $\mathbb H$ \cite{lokhin1963nonlinear, boehler1977irreducible, boehler1979simple}; i.e., 
\begin{equation}
\bar \psi (\mathbf E^{(0)}) = \hat \psi(\mathcal J_1, \mathcal J_2, ..., \mathcal J_n), 
\end{equation}
where
\begin{equation}
       {\mathcal J}_i(\mathbf{E}^{(0)}; \, \mathbb H) =  {\mathcal J}_i(\mathbf{Q}^T\mathbf{E}^{(0)}\mathbf{Q}; \, {\mathbb P}_{\mathbf{Q}}(\mathbb H))
      \, \forall \mathbf{Q} \in SO_2.
      \end{equation}
      Here  $ {\mathbb P}_{\mathbf{Q}}$ denotes the  transformation of the structure tensor $\mathbb H$ under the orthogonal transformation $\mathbf{Q}$.
These special functions are termed symmetry invariants since they satisfy all the symmetry constraints belonging to the material symmetry group of the crystal. Smith \cite{Smith1, Smith2, Smith3} showed that the set of mutually-independent symmetry invariants --- which is finite for all cases of material symmetry --- serves as a  complete and irreducible basis for the representation of scalar constitutive functions of the anisotropic material. In the following section, we explicitly derive the symmetry invariants of $\mathbf E^{(0)}$ for the structure tensor characterizing the material symmetry group of  graphene.
\section{Hyperelastic constitutive response of graphene}
\label{hyper}
Following the invariant-theoretic approach outlined in Sec.~\ref{kine},  we now systematically construct a hyperelastic constitutive response function for arbitrary in-plane deformation of graphene. First, we explicitly obtain the structure tensor characterizing the material symmetry group of graphene. The structure tensor $\mathbb H$ corresponding to a $\mathcal C_{(2n)v}$ material symmetry group is obtained by a general expression given by \cite{Zheng1, Zheng2, Zheng3}
\begin{equation}
\mathbb H = \text{Re} (\mathbf M + i \mathbf N)^{\otimes n},
\label{genexp}
\end{equation}
where $(...)^{\otimes n} = (...) \otimes (...)  \otimes ... \otimes (...)  (n \text{ times})$, and $2n$ denotes the order of the principal rotation axis. $\mathbf M$ and $\mathbf N$ are dimensionless symmetric traceless tensors given by
\begin{equation}
\mathbf M =  \hat{\mathbf x} \otimes\hat{\mathbf x} - \hat{\mathbf y} \otimes\hat{\mathbf y}; \quad \quad 
\mathbf N =  \hat{\mathbf x} \otimes\hat{\mathbf y} + \hat{\mathbf y} \otimes\hat{\mathbf x}, 
\end{equation}
where $\hat{\mathbf x}$ and $\hat{\mathbf y}$ denote  two orthogonal material unit vectors fixed in the frame of the  reference crystal such that at least one of them is aligned with an axis of reflection symmetry.\\
From Eq.~(\ref{genexp}), we obtain the sixth-order structure tensor characterizing $\mathcal C_{6v} (n=3)$, the material symmetry group of graphene, as 

\begin{equation}
\mathbb H = \mathbf M \otimes \mathbf M \otimes \mathbf M  - \left(\mathbf M \otimes \mathbf N \otimes \mathbf N +\mathbf N \otimes \mathbf M \otimes \mathbf N + \mathbf N \otimes \mathbf N \otimes \mathbf M \right).
\end{equation}
\begin{minipage}{\linewidth}
\centering
\includegraphics[scale=0.5]{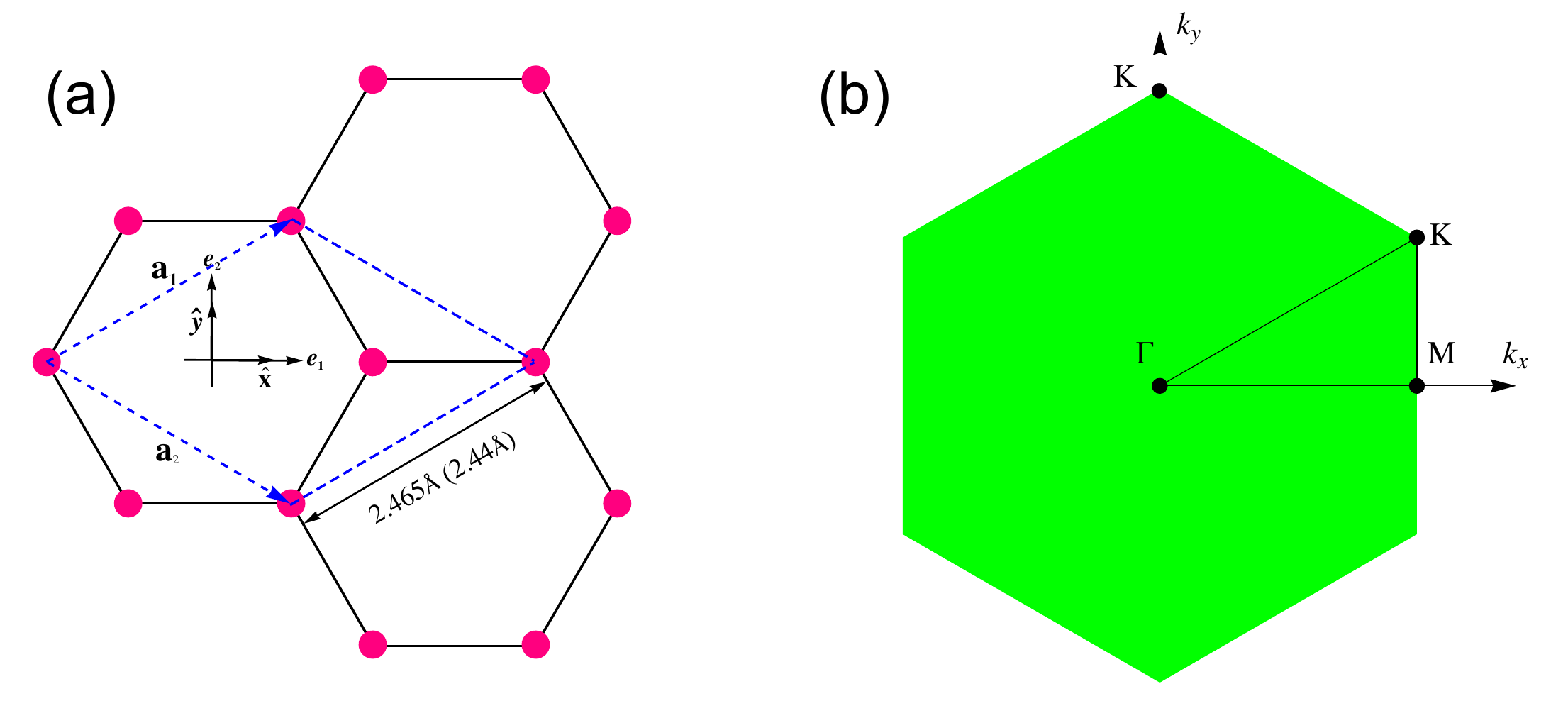}
\captionof{figure}{ (a) Graphene lattice with orientations of the material unit vectors --- $\hat{\mathbf x}$ and $\hat{\mathbf y}$ --- and the Cartesian unit vectors --- $\mathbf e_1$ and $\mathbf e_2$ --- indicated. The dashed lines denote the unit cell used in the \textit{ab initio} calculations. The GGA (LDA) value of the lattice parameter is also indicated. The armchair and zigzag directions are along the $\mathbf e_1$ and $\mathbf e_2$ axes respectively. (b) Brillouin zone of graphene with high symmetry points indicated. }
\label{lattice}
\end{minipage}
\\
The complete and irreducible set of polynomial joint invariants of $\mathbf E^{(0)}$ and $\mathbb H$ constitute the symmetry invariants of $\mathbf E^{(0)}$.  Following the procedure of  Zheng and Betten \cite{Zheng1}, we obtain  three independent scalar joint invariants of this 2D system as
 \begin{equation}
 \mathcal J_1 \equiv \epsilon_a= \mathrm{tr} \mathbf E^{(0)}  = \ln J,
 \end{equation}
 \begin{equation}
  \mathcal J_2 \equiv   (\gamma_i/2)^2= \frac{1}{2}\ \mathbf E^{(0)}_0 : \mathbf E^{(0)}_0 =  (\ln \lambda)^2,
  \end{equation}
  where $\mathbf A : \mathbf B = \mathrm{tr}(\mathbf A^{T} \mathbf B)$ is the scalar tensor product,
  and\\
  \begin{multline}
  \mathcal J_3 \equiv (\gamma_{\theta}/2)^3= \frac{1}{8}\mathbb H[\mathbf E^{(0)}_0,\mathbf E^{(0)}_0,\mathbf E^{(0)}_0]\\
  = \frac{1}{8}\left[ \left(\mathbf{M} : \mathbf{E}_0^{(0)}\right)^3 -3\left(\mathbf{M} :\mathbf{E}_0^{(0)}\right) 
  \left(\mathbf{N} : \mathbf{E}_0^{(0)}\right)^2 \right] \\
  = \left(\ln \lambda \right)^3 \, \cos 6 \theta,
  \end{multline}
where $\cos \theta = \mathbf r_1 . \hat{\mathbf x}$ indicates the orientation of maximum principal stretch. 
The first two of these invariants, $\epsilon_a$ and $\gamma_i \ge 0$, are simply two isotropic invariants of $\mathbf E^{(0)}$ alone. Thus any material anisotropy in the constitutive response of graphene is captured solely  by the third invariant $\gamma_{\theta}$. \\
  In order to  represent second-order  tensorial quantities, we need the following list of form-invariants:
 \begin{equation}
 \frac{\partial \epsilon_a}{\partial \mathbf E^{(0)}} = \mathbf I;
 \end{equation}
 \begin{equation} 
 \frac{\partial (\gamma_i^2)}{\partial \mathbf E^{(0)}} = 4\mathbf E^{(0)}_0;
 \end{equation}
 \begin{equation}
  \frac{\partial (\gamma_{\theta}^3)}{\partial \mathbf E^{(0)}} = 3 \bigg \lgroup \left[ (\mathbf M : \mathbf E^{(0)}_0)^2 -(\mathbf N : \mathbf E^{(0)}_0)^2 \right]  \mathbf M - \left[2(\mathbf M : \mathbf E^{(0)}_0)(\mathbf N : \mathbf E^{(0)}_0) \right]  \mathbf N \bigg \rgroup \equiv \mathbf S_{\mathbf E^{(0)}_0}.
  \end{equation}
 For purposes of constructing  fourth-order tangent modulus tensors,  second derivatives of the invariants with respect to $\mathbf E^{(0)}$ are obtained as follows:
  \begin{equation}
   \frac{\partial^2 \epsilon_a}{\partial \mathbf E^{(0)} \partial \mathbf E^{(0)}} =\mathbb O,
   \end{equation}
   \begin{equation}
 \frac{\partial^2 (\gamma_i^2)}{\partial \mathbf E^{(0)} \partial \mathbf E^{(0)}} = 4 \left[\mathbb I - \frac{1}{2} \mathbf I \otimes \mathbf I \right],
 \end{equation}
 \begin{equation}
  \frac{\partial^2 (\gamma_{\theta}^3)}{\partial \mathbf E^{(0)}\partial \mathbf E^{(0)}} = 6 \bigg \lgroup  \left(\mathbf M : \mathbf E^{(0)}_0 \right) \,\mathbf M\otimes \mathbf M -\left(\mathbf N : \mathbf E^{(0)}_0\right) \, \left[\mathbf M\otimes \mathbf N + \mathbf N\otimes \mathbf M + \mathbf N \otimes \mathbf N \right]\bigg \rgroup \equiv \mathbb S_{\mathbf E^{(0)}_0}.
 \end{equation}
 Here $\mathbb O$ and $\mathbb I$ are the fourth-order zero and identity tensors, respectively.
\subsection{Strain energy per unit reference area $\psi$}
\label{rep}
 The proposed hyperelastic model is based on representation of the strain energy per unit reference area $\psi$ in terms of the symmetry invariants of $\mathbf E^{(0)}$, i.e., $\psi=  \hat \psi(\epsilon_a, \gamma_i, \gamma_{\theta})$. 
 For application of invariant-based hyperelasticty,  the  strain measure $\mathbf E^{(0)}$  offers substantial simplifications in terms of formulation. First, as is evident from its spectral representation (Eq.~(\ref{log})), $\mathbf E^{(0)}$ additively decomposes  areal ($\mathbf U^a$) and  isochoric ($\tilde {\mathbf U}$) parts of the deformation into isotropic and deviatoric parts of the strain, respectively. Secondly, the state nature of $\psi$ in a hyperelastic material enables calculating $\psi$ by integrating $d \psi = (\partial \psi/\partial \mathbf E^{(0}): d \, \mathbf E^{(0)}$ along any convenient strain path, where it is understood that $\mathbf T^{(0)} \equiv \partial \psi/ \partial \mathbf E^{(0)}$ is the work-conjugate stress measure. Let a first, purely isotropic strain path (Path 1) correspond to  areal deformation $\mathbf U^{a}$ while holding $\tilde{\mathbf U} = \mathbf I$, and let a second, purely deviatoric strain path (Path 2) correspond to a subsequently-imposed isochoric deformation $\tilde{\mathbf U}$ while holding $\mathbf U^a = J^{1/2} \,\mathbf I$.   The isotropic/deviatoric decomposition of the work-conjugate stress is $\mathbf T^{(0)}  \equiv \frac{1}{2} S^{(0)} \mathbf I + \mathbf T^{(0)}_0$,
  while that of the incremental log strain is $ d \mathbf E^{(0)}  = \frac{1}{2} \, d \epsilon^a \, \mathbf I + d \mathbf E^{(0)}_0$; thus, the
  incremental energy/work relation is additively decomposed into isotropic and deviatoric parts
 \begin{equation}
 d \, \psi = \mathbf T^{(0)}: \, d \mathbf E^{(0)} = \frac{1}{2} S^{(0)} \,d \epsilon^a + \mathbf{T}^{(0)}_0 : d \mathbf E^{(0)}_0.
 \end{equation}
 The strain energy of an arbitrary deformed state of the lattice --- when expressed in terms of invariants of the log strain --- can be additively decomposed into a term $\psi^{\mathrm{Dil}} (\epsilon_a)$ --- corresponding to a pure areal deformation $\mathbf U^a$ while $\mathbf E^{(0)}_0 =\mathbf{0}$ --- plus  a term $\psi^{\mathrm{Dev}} (\gamma_i, \gamma_{\theta}; \epsilon_a)$ --- corresponding to a superimposed  isochoric deformation $\tilde{\mathbf U} $ while $\text{tr}\,\mathbf E^{(0)}$ remains fixed at $\epsilon_a$.  Thus, we write:
\begin{equation}
\psi=  \hat \psi(\epsilon_a, \gamma_i, \gamma_{\theta})= \hat \psi^{\mathrm{Dil}}(\epsilon_a) +\hat \psi^{\mathrm{Dev}} ( \gamma_i, \gamma_{\theta}; \epsilon_a).
\label{uber}
\end{equation}
 Setting $\psi=0$ in the undeformed configuration, the areal contribution $\psi^{\mathrm{Dil}}$ equals the isotropic stress working  along Path 1, along which  $d \mathbf E^{(0)}_0 = \mathbf 0$; thus 
 $\psi^{\mathrm{Dil}} = \hat \psi^{\mathrm{Dil}} (\epsilon_a)$.  The contribution $\psi^{\mathrm{Dev}}$  is numerically equal to the
 deviatoric stress working along the subsequent strain Path 2, along which $d \epsilon^a = 0$; thus $\psi^{\mathrm{Dev}}$  depends on invariants $\gamma_i$ and $\gamma_{\theta}$ of the  imposed deviatoric strain $\mathbf{E}^{(0)}_0$, as well as having  implicit dependence on the  (constant) areal strain $\epsilon_a$ along Path 2, with `initial' condition $\psi^{\mathrm{Dev}}( 0, 0; \epsilon^a) = 0$.  \\
 
  From elasticity theory, we recall that the flexural stiffness $D$ of a thin structural element scales with its thickness $h$ as $D \sim h^3$, whereas the stretching stiffness $C$ scales linearly with $\sim h$. Notably, graphene is just one atomic layer thick,  $h \approx 10^{-10}$m, so in a superthin structure like graphene, the ratio of bending stiffness to in-plane stiffness is exceedingly small and, accordingly, we assume that the contribution of bending to the strain energy per unit area is negligible compared to that of in-plane strain.
  Furthermore, and for the same reason, a suspended graphene sheet under a compressive in-plane loading, i.e., a state of Cauchy stress $\pmb \sigma$ with $\mathbf n. \pmb \sigma. \mathbf n <0 $ for some in-plane direction $\mathbf n$, is structurally unstable in the limiting case of vanishing flexural rigidity, and will immediately buckle out-of-plane. Therefore, the scope of the modeling effort is limited to only those in-plane deformation states for which $\mathbf n. \pmb \sigma. \mathbf n \ge 0 
  \, \forall \, \mathbf n$.   \\
 
 \textbf{ (a). Energetic response under pure dilation, $\pmb \psi^{\textbf {Dil}}$ :}
The energetic response under pure dilation, i.e.,  when $\mathbf U = {\mathbf U}^{a} = J^{1/2} \, \mathbf I$ , is well-described by a function based on the universal binding energy relation (UBER) proposed by Rose  \textit{et al.} \cite{rose1983universal, rose1981universal}.  The UBER relation is
\begin{equation}
\hat \psi^{\mathrm{Dil}}(\epsilon_a) = \mathcal E\left[1 - (1+ \alpha \epsilon_a) \exp (-\alpha \epsilon_a)\right].
\label{psidil}
\end{equation}
 Table~\ref{Tabcoeffs} lists values of the constants $\mathcal E$ and  $\alpha$ as determined by fitting the \textit{ab initio} energies, calculated using both LDA and GGA formulations, for the graphene lattice subjected to pure dilatory deformations. 
\begin{table}[htbp!]
\begin{center}
\begin{tabular}{c  c    c  c  c}
\toprule[1.5pt] \\
&& $\alpha$ \hspace*{1.5cm} &   $\mathcal E $(N/m) & $\epsilon_a\rvert_{\kappa=0} = 1/(1+\alpha)$ \\
\hline 
\hline
\\
GGA \hspace*{1.5cm}& &1.53 \hspace*{0.5cm}  & 93.84 & 0.40 \\ \\
\hline
\\
LDA \hspace*{1.5cm}&& 1.38 \hspace*{0.5cm}  & 116.43 & 0.42 \\ \\
\bottomrule[1.5pt]
\end{tabular}
\end{center}
\caption{ Values of $\mathcal E$ and $\alpha$ obtained by fitting the UBER form (Eq.~(\ref{psidil})) to \textit{ab initio} energies. The UBER constitutive form exhibits a tangent area modulus $\kappa$ ---defined in  Eq.~(\ref{bm})--- that decreases with increasing areal strain, ultimately vanishing at a critical areal strain value  given by $\epsilon_a\rvert_{\kappa=0}=1/(1+\alpha)$. }
\label{Tabcoeffs}
\end{table}\\
As shown in Fig.~[\ref{psi}-(a)], the UBER-based model for the fitted values of $\mathcal E$ and $\alpha$  accurately describes the \textit{ab initio} energies for pure areal change. \\ \\ 

\begin{minipage}{\linewidth}
\includegraphics[scale=0.6]{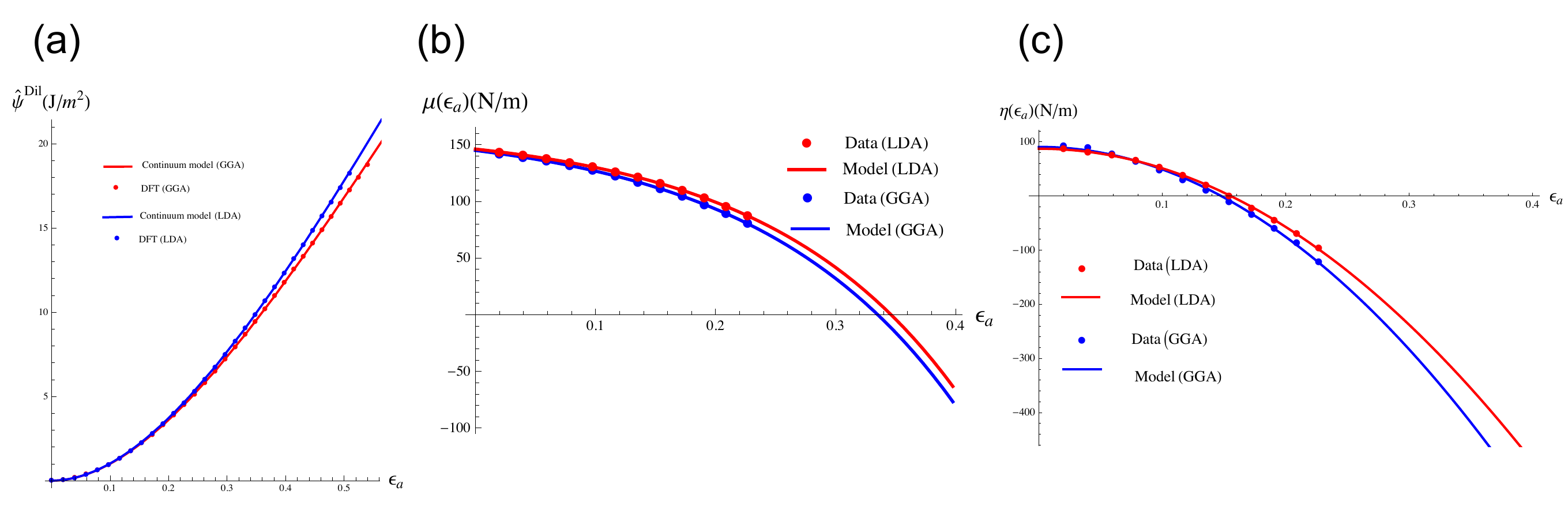}
\captionof{figure}{(a) The universal binding energy relation-based model for volumetric response $\hat \psi^{\mathrm{Dil}}$ for fitted values of the coefficients $\mathcal E$ and $\alpha$ as tabulated in Table~\ref{Tabcoeffs}. For comparison, we  also show the \textit{ab initio} energies used in fitting the UBER form. (b) DFT data for $\mu(\epsilon_a)$ fitted with the proposed functional form of Eq.~(\ref{mue}). (c) DFT data for $\eta(\epsilon_a)$ fitted with the proposed functional form of Eq.~(\ref{nue}). }
\label{psi}
\end{minipage}
\\ \\

\textbf{(b). Energetic response under shape-changing deformations, $\pmb \psi^{\textbf{Dev}}$: }
The shear stiffness of graphene,  in general, is dilation-sensitive, depending upon $\epsilon_a$. Thus, in general, we assume that  $\psi^{\mathrm{Dev}} =\hat \psi^{\mathrm{Dev}}(\epsilon_a, \gamma_i, \gamma_{\theta} )$. \\ 

We find that a simple linear combination of monomials in $ \mathcal J_2 = (\gamma_i/2)^2$ , and in $\mathcal J_3=(\gamma_{\theta}/2)^3$, with coefficients that are functions of $\epsilon_a$;  i.e.,
\begin{equation*}
\hat \psi^{\mathrm{Dev}}(\gamma_i, \gamma_{\theta}; \epsilon_a) = \frac{1}{2} \mu (\epsilon_a) \gamma_i^2 +\frac{1}{8} \eta(\epsilon_a) \gamma_{\theta}^3,
\end{equation*}
fits the \textit{ab initio} calculations well using simple functional forms for $\mu(\epsilon_a)$ and $\eta(\epsilon_a)$.
The shear modulus $\mu$ is well-fit by an exponentially-decreasing function of the areal strain, $\epsilon_a$ (see Fig.~[\ref{psi}-(b)]) ,
\begin{equation}\mu(\epsilon_a) = \mu_0 - \mu_1 e^{\beta \epsilon_a},
\label{mue}
\end{equation}
whereas $\eta(\epsilon_a)$ is fit by an even quadratic function of $\epsilon_a$ (Fig.~[\ref{psi}-(c)]), 
\begin{equation}
\eta(\epsilon_a) = \eta_0 - \eta_1 \epsilon_a^2.
\label{nue}
 \end{equation}
 Thus, the overall expression  proposed for the free energy $\psi$ is
 \begin{equation}
 \hat \psi =  \mathcal E -\mathcal E (1+ \alpha \epsilon_a) \exp (-\alpha \epsilon_a) + \frac{1}{2}(\mu_0 - \mu_1 e^{\beta \epsilon_a} ) \gamma_i^2 +\frac{1}{8} ( \eta_0  - \eta_1 \epsilon_a^2) \gamma_{\theta}^3.
\end{equation}
The constants in the expressions for $\mu$ and $\eta$ are determined by fitting the set of  \textit{ab initio} energies  calculated for a number of deformed states described in Sec.~\ref{train}; fitted values are summarized in Tables~[\ref{mucoeff}] and [\ref{nucoeff}].\\
Using a total of only 7 scalar fitting parameters,  half the number used in the formulation of Wei  \textit{et al.}, the proposed functional form fits the entire DFT dataset very well.\\

\begin{table}[htbp!]
\begin{center}
\begin{tabular}{c c c c c }
\toprule[1.5pt] 

&$\mu_0 $(N/m) & $\mu_1$(N/m) & $\beta$& $\epsilon_a\rvert_{\mu =0}=\frac{1}{\beta}\ln(\mu_0/\mu_1)$ \\
\hline 
\hline
 & & & \\
GGA \hspace*{1cm} & 172.18  & 27.03 & 5.32 & 0.35\\
\hline
\\
LDA \hspace*{1cm}& 164.17 & 17.31 & 6.32  & 0.36 \\
\bottomrule[1.65pt]
\end{tabular}
\end{center}
\caption{Coefficients $\mu_0$ and $\mu_1$ and exponent $\beta$ in Eq.~(\ref{mue}) for  shear modulus $\mu(\epsilon_a)$, determined by  least-square fits to a set of \textit{ab initio} energies. The data and the functional form suggest that $\mu$ vanishes at critical value of areal strain given by $\epsilon_a\rvert_{\mu =0}=\frac{1}{\beta}\ln(\mu_0/\mu_1)$, tabulated in the right-most column.}
\label{mucoeff}
\end{table}

\begin{table}[htbp!]
\begin{center}
\begin{tabular}{c c c }
\toprule[1.5pt]
&$\eta_0 $(N/m) & $\eta_1$(N/m) \\
\hline 
\hline
 & & \\
GGA \hspace*{1cm}& 94.65  \hspace*{1cm} & 4393.26 \\
\hline
\\
LDA \hspace*{1cm} & 93.17  \hspace*{1cm}& 4408.76 \\
\bottomrule[1.65pt]
\end{tabular}
\end{center}
\caption{Coefficients $\eta_0$ and $\eta_1$ in Eq.~(\ref{nue}) for $\eta(\epsilon_a)$,  determined by
least-squares fitting to a set of \textit{ab initio} energies.}
\label{nucoeff}
\end{table}
\subsection{Work-conjugate stress tensor $\mathbf T^{(0)}$}
\label{wconj}
Using the proposed functional forms for $\psi$, the stress measure $\mathbf T^{(0)}$  work-conjugate to $\mathbf{E}^{(0)}$ is calculated as
\begin{widetext}
\begin{multline}
\mathbf T^{(0)}(\epsilon_a, \gamma_i, \gamma_{\theta})= \frac{\partial \psi }{\partial \mathbf E^{(0)}} \\
= \bigg \lgroup \frac{\partial \hat \psi^{\mathrm{Dil}}(\epsilon_a)}{ \partial \epsilon_a} +\frac{\partial \hat \psi^{\mathrm{Dev}}(\epsilon_a, \gamma_i, \gamma_{\theta})}{ \partial \epsilon_a}\bigg \rgroup \mathbf I   + 4 \frac{\partial \hat \psi^{\mathrm{Dev}}(\epsilon_a, \gamma_i, \gamma_{\theta})}{ \partial (\gamma_i^2)}  \mathbf E^{(0)}_0 + \frac{\partial \hat \psi^{\mathrm{Dev}}(\epsilon_a, \gamma_i, \gamma_{\theta})}{ \partial (\gamma_{\theta}^3)}\mathbf S_{\mathbf E^{(0)}_0} \\
=\bigg \lgroup \mathcal E \alpha^2 \epsilon_a \exp(-\alpha\epsilon_a)+ \frac{1}{2}\mu^{\prime}(\epsilon_a)\gamma_i^2+ \frac{1}{8} \eta^{\prime}(\epsilon_a) \gamma_{\theta}^3\bigg \rgroup \mathbf I + 2 \mu(\epsilon_a) \mathbf E^{(0)}_0 + 
\frac{1}{8}\eta(\epsilon_a)\mathbf S_{\mathbf E^{(0)}_0},
\label{wcstress}
\end{multline}
\end{widetext}
where the prime notation $(\ldots)^{\prime}$ denotes differentiation with respect to $\epsilon_a$.
Following the hyperelastic work-conjugacy relation (see Love \cite{love2013treatise}, Ogden\cite{ogden1997non}), the conversion from work-conjugate stress $\mathbf T^{(0)}$ to Cauchy stress $\pmb \sigma$ is obtained as follows. Letting a superposed dot denote the material time derivative, the 
power balance of isothermal hyperelasticity identifies measures of stress that are power-conjugate to differing
measures of strain-rate by
\begin{equation}
\dot{\psi} = \mathbf T^{(2)}: \dot{\mathbf E}^{(2)} = \mathbf T^{(0)} :\dot{\mathbf E}^{(0)},
\label{wc}
\end{equation}
etc., where, for example, $\mathbf E^{(2)} = \frac{1}{2} (\mathbf C -\mathbf I) =  \frac{1}{2} (\mathbf F^T \mathbf F -\mathbf I)$ is the Green-Lagrange strain tensor, and $\mathbf T^{(2)}$ is its power-conjugate stress tensor, often denoted as the second Piola-Kirchhoff stress tensor. Using the chain rule, and viewing $\mathbf E^{(0)}$ as a function of $\mathbf E^{(2)}$, we obtain
\begin{equation}
\dot{\mathbf E}^{(0)} =\left[ \frac{\partial \mathbf E^{(0)} }{\partial \mathbf E^{(2)} }\right] \dot{\mathbf E}^{(2)} = 2 \underbrace{\left[ \frac{\partial \mathbf E^{(0)} }{\partial \mathbf C \ \ \  }\right]}_{\large \equiv {\mathcal L^{(1)}}} \dot{\mathbf E}^{(2)}.
\label{dedc}
\end{equation}
where the fourth-order tensor ${\mathcal L^{(1)}} \equiv \frac{\partial \mathbf E^{(0)} }{\partial \mathbf C \ \ \  } $ has major symmetry, i.e., ${\mathcal L^{(1)}}^T = {\mathcal L^{(1)}}$, and  it is understood that  $\mathcal L^{(1)}$ operates on the second-order tensor appearing on its immediate right to produce a resultant second-order tensor.
Substituting Eq.~(\ref{dedc}) into Eq.~({\ref{wc}}) gives
\begin{equation}
\mathbf T^{(2)}: \dot{\mathbf E}^{(2)}=\mathbf T^{(0)} :  \bigg \lgroup 2 \mathcal L^{(1)}\dot{\mathbf E}^{(2)}\bigg \rgroup = \bigg \lgroup 2 {\mathcal L^{(1)} }^T\mathbf T^{(0)}\bigg \rgroup : \dot{\mathbf E}^{(2)} = \bigg \lgroup 2 {\mathcal L^{(1)} }\mathbf T^{(0)}\bigg \rgroup : \dot{\mathbf E}^{(2)}.
\label{conv}
\end{equation}
In order for Eq.~(\ref{conv}) to hold for arbitrary $\dot{\mathbf E}^{(2)}$, we obtain the transformation relation
\begin{equation}
\mathbf T^{(2)} = 2 \mathcal L^{(1)} \mathbf T^{(0)}.
\end{equation}
The Cauchy stress $\pmb \sigma$ can be similarly-obtained from power balance relations  (see Ogden \cite{ogden1997non}) as
\begin{equation}
\pmb \sigma = \frac{1}{J}\,  \mathbf F \mathbf T^{(2)}\mathbf F^T = \frac{1}{J} \, \mathbf F \bigg \lgroup 2 \mathcal L^{(1)} \mathbf T^{(0)} \bigg \rgroup \mathbf F^T .
\end{equation}
Evaluation of $\mathcal L^{(1)}$ in a general state of deformation is carried out in Sec.~\ref{acou}.\\
As a particular illustration, we can use Eq.~(\ref{wcstress}) to evaluate $\mathbf{T}^{(0)}$ for equi-biaxial stretch. In the absence of shape change,  $ \mathbf{E}_0^{(0)} = \mathbf{0}$ and hence $\gamma_i=\gamma_{\theta}=0$, so that 
\begin{equation}
\mathbf T^{(0)}(\epsilon_a)= \frac{\partial \hat \psi^{\mathrm{Dil}}}{\partial \epsilon_a} =  \mathcal E \alpha^2 \epsilon_a e^{-\alpha \epsilon_a}\mathbf I.
\end{equation}
Further, in this special deformation case, $\mathcal L^{(1)}$ can be shown to have the simple form:
\begin{equation}
\mathcal L^{(1)} = \frac{\partial \mathbf E^{(0)} }{\partial \mathbf C \ \ \ } = \frac{1}{2J} \, \mathbb I,
\end{equation}
so that $\pmb \sigma$ under equi-biaxial stretch becomes
\begin{equation}
\pmb \sigma \rvert_{\mathbf{F} = J^{1/2}\,\mathbf{I} }= 2 \, \frac{1}{2J}\,  \mathbf I\bigg \lgroup \mathbb I \mathbf T^{(0)}\rvert_{\mathbf{F} = J^{1/2}\,\mathbf{I}}\bigg \rgroup \mathbf I = \frac{1}{J} \, \mathbf T^{(0)} \rvert_{\mathbf{F} = J^{1/2}\,\mathbf{I}};
\end{equation}
 thus, the equi-biaxial Cauchy stress is
\begin{equation}
\pmb\sigma \rvert_{\mathbf{F} = J^{1/2} \mathbf I}  = \frac{1}{J} \, \frac{\partial \hat \psi^{\mathrm{Dil}}}{\partial \epsilon_a}\, \mathbf I= \mathcal E \alpha^2 \epsilon_a e^{-(1+\alpha) \epsilon_a}\,  \mathbf I \equiv p \, \mathbf I,
\end{equation}
where $p \equiv \frac{1}{2} \mathrm{tr} \, \pmb \sigma $, 
and a corresponding tangent areal modulus $\kappa$ is
\begin{equation}
\kappa(\epsilon_a) \equiv \frac{d p(\epsilon_a)}{d \epsilon_a} =  \mathcal E \, \alpha^2  \Big(1-\left(1+\alpha\right)\,\epsilon_a \Big)e^{-\left(1+\alpha\right) \epsilon_a}.
\label{bm}
\end{equation}
Under equi-biaxial stretch, the equi-biaxial Cauchy stress reaches its maximum when 
\begin{equation}
\kappa = 0 \Rightarrow1-\left(1+\alpha\right)\epsilon_a=0.
\end{equation}
Thus, the peak equi-biaxial Cauchy stress occurs at a critical areal strain given by (see also  Table~(\ref{coeffs})):
\begin{equation}
\epsilon_a\rvert_{\kappa=0} = \frac{1}{1+\alpha}.
\end{equation}
\subsection{Work-conjugate tangent moduli tensor $\mathbb L^{(0)}$}
\label{tang}
The fourth-order tensor of tangent moduli connecting $\dot{\mathbf T}^{(0)}$ to $\dot{\mathbf E}^{(0)}$ is defined as
\begin{equation}
\mathbb L^{(0)} = \frac{\partial \mathbf T^{(0)}}{\partial \mathbf E^{(0)}} = \frac{\partial^2 \, \psi}
{\partial \mathbf{E}^{(0)} \, \partial \mathbf{E}^{(0)}}.
\end{equation}
On using Eq.~(\ref{wcstress}), $\mathbb L^{(0)}$ can be straightforwardly expressed as
\begin{widetext}
\begin{multline}
\mathbb L^{(0)} =\bigg \lgroup \mathcal E \alpha^2 (1-\alpha\epsilon_a) \exp(-\alpha\epsilon_a)+ \frac{1}{2}\mu^{\prime \prime}(\epsilon_a)\gamma_i^2+ \frac{1}{8} \eta^{\prime \prime}(\epsilon_a) \gamma_{\theta}^3\bigg \rgroup \mathbf I \otimes \mathbf I+ 2 \mu^{\prime}(\epsilon_a)\bigg \lgroup \mathbf I \otimes \mathbf E^{(0)}_0 + \mathbf E^{(0)}_0 \otimes \mathbf I \bigg \rgroup \\ + \frac{1}{8} \eta^{\prime}(\epsilon_a)\,
\bigg \lgroup \mathbf I \otimes \mathbf S_{\mathbf E^{(0)}_0}+\mathbf S_{\mathbf E^{(0)}_0}\otimes \mathbf I \bigg \rgroup+
2 \mu(\epsilon_a)\, \bigg \lgroup \mathbb I - \frac{1}{2} \mathbf I \otimes \mathbf I \bigg \rgroup + \frac{1}{8}\eta(\epsilon_a) \,  \mathbb S_{\mathbf E^{(0)}_0}.
\end{multline}
\end{widetext}
\subsection{Acoustic tensor}
\label{acou}
The acoustic tensor is defined as the second derivative of the strain energy density function with respect to the deformation gradient tensor $\mathbf F$ \cite{Born1,wallace1998thermodynamics,ogden1997non}:
\begin{equation}
\mathbb A = \frac{\partial^2 \psi}{ \partial \mathbf F^2}.
\end{equation}
 The first derivative of the free energy $\psi$ with respect to $\mathbf F$ gives the generally non-symmetric first Piola-Kirchhoff stress tensor, $\mathbf T^{PK1}$, which can be evaluated using the chain rule as:
\begin{equation}
\left[\frac{\partial \psi (\mathbf E^{(0)})}{\partial \mathbf F}\right]_{ij}= \left[ \mathbf T^{PK1} \right]_{ij}= \underbrace{\left[\frac{\partial \psi (\mathbf E^{(0)})}{\partial \mathbf E^{(0)}}\right]_{mn}}_{\text {Work-conjugate stress,} \mathbf T^{(0)}_{mn}}  \underbrace{\frac{\partial \mathbf E^{(0)}_{mn}}{\partial \mathbf C_{pq}} \frac{\partial \mathbf C_{pq}}{\partial \mathbf F_{ij}}}_{\partial \mathbf E^{(0)}_{mn}/ \partial \mathbf F_{ij}} .
\end{equation}
Then the second derivative, which is the acoustic tensor, is obtained as:
\begin{multline}
\mathbb A _{ijkl} = \left[\frac{\partial^2 \psi (\mathbf E^{(0)})}{\partial \mathbf F^2}\right]_{ijkl}=
 [\mathbb L^{(0)}]_{mnrs} \left[\frac{\partial \mathbf E^{(0)}}{\partial \mathbf C} \right]_{mnpq} \left[ \frac{\partial \mathbf C}{\partial \mathbf F} \right]_{pqij}\left[ \frac{\partial \mathbf E^{(0)}}{\partial \mathbf C}\right]_{rsab} \left[ \frac{\partial \mathbf C}{\partial \mathbf F}\right]_{abkl} \\ \hspace*{5cm} +[\mathbf T^0]_{mn} \left[\frac{\partial^2 \mathbf E^{(0)}}{\partial \mathbf C \partial \mathbf C} \right]_{mnpqrs} \left[\frac{\partial \mathbf C}{\partial \mathbf F}\right]_{rskl} \left[\frac{\partial \mathbf C}{\partial \mathbf F}\right]_{pqij}\\ 
 +[\mathbf T^0]_{mn} \left[\frac{\partial \mathbf E^{(0)}}{\partial \mathbf C}\right]_{mnpq} \left[\frac{\partial^2 \mathbf C}{\partial \mathbf F \partial \mathbf F}\right]_{pqijkl}.
 \label{acoustictensor}
\end{multline}
In the present applications, the acoustic tensor $\mathbb A$ is a fourth-order tensor in 2D. By definition, it has the major symmetry  $\mathbb A_{ijkl} = \mathbb A_{klij}$.  The representation of the  fourth-order tensor with respect to the orthonormal basis $\{\mathbf e_i \otimes \mathbf e_j \otimes \mathbf e_k \otimes \mathbf e_l \}_{1 \le i,\ j,\ k, \ l \le 2}$ is given by
\begin{equation}
\mathbb A = \mathbb A_{ijkl}\  \mathbf e_i \otimes \mathbf e_j \otimes \mathbf e_k \otimes \mathbf e_l,
\end{equation}
where  $\mathbb A_{ijkl} = (\mathbf e_i \otimes \mathbf e_j): \mathbb A: (\mathbf e_k \otimes \mathbf e_l )$, with $1 \le i,\ j,\ k, \ l \le 2$. We can define another orthonormal basis set \cite{moakher2008fourth}  $ \{ \tilde{\mathbf e}_{\alpha} \otimes \tilde{\mathbf e}_{\beta}  \}_{1 \le \alpha,\ \beta \le 4}$ where $\tilde{\mathbf e}_{f(i,j)} = \mathbf e_i \otimes \mathbf e_j$, with $f(1,1) =1$,  $f(2,2) =2$, $f(1,2) =3$, and $f(2,1) =4$. In this alternate basis, $\mathbb A$ can be represented by a 4-dimensional second-order tensor $\tilde{\mathbf A}$ given by
\begin{equation}
 \tilde{\mathbf A} = \tilde{\mathbf A}_{\alpha \beta} \  \tilde{\mathbf e}_{\alpha} \otimes \tilde{\mathbf e}_{\beta}.
\end{equation}
We can thus express the acoustic tensor $\mathbb A$ in terms of a $4 \times 4$ matrix of the following form
\begin{equation}
\mathbb A = \begin{bmatrix}
\mathbb A_{1111} & \mathbb A_{1122} & \mathbb A_{1112} & \mathbb A_{1121} \\
\mathbb A_{1122} & \mathbb A_{2222} & \mathbb A_{2212} & \mathbb A_{2221} \\
\mathbb A_{1112} & \mathbb A_{2212} & \mathbb A_{1212} & \mathbb A_{1221} \\
\mathbb A_{1121} & \mathbb A_{2221} & \mathbb A_{1221} & \mathbb A_{2121} \\
\end{bmatrix}.
\end{equation}
The tensor derivatives appearing in the expressions for $\mathbf T^{(0)}$, $\mathbb L^{(0)}$, and $\mathbb A$  are calculated following the methods prescribed by Ogden \cite{ogden1997non} , Norris \cite{norris2007eulerian} 
and Carlson and Hoger \cite{carlson1986derivative}.
\begin{enumerate}
\item \begin{equation}
\left[\frac{\partial \mathbf C }{\partial \mathbf F}\right]_{ijkl} =\frac{\partial \mathbf C_{ij} }{\partial \mathbf F_{kl}}=  \delta_{il} [\mathbf F]_{kj}+ [\mathbf F]_{ki} \delta_{jl}.
\end{equation}
\item \begin{equation}
\left[ \frac{\partial^2\mathbf C}{\partial \mathbf F \partial \mathbf F}\right]_{ijklmn}= \frac{\partial^2\mathbf C_{ij}}{\partial \mathbf F_{kl} \partial \mathbf F_{mn}}= \delta_{il} \delta_{km} \delta_{jn}+ \delta_{km} \delta_{in} \delta_{jl}.
\end{equation}
\item Following Ogden \cite{ogden1997non} and Norris \cite{norris2007eulerian}, based on the principal axes technique introduced by Hill \cite{hill1970constitutive}, we obtain
\begin{eqnarray}
\mathcal L^{(1)} = \frac{\partial \mathbf E^{(0)} }{\partial \mathbf C}= \begin{cases}
&\\
\frac{1}{2\lambda_1^2} \mathbf U_1 \boxtimes \mathbf U_1+ \frac{1}{2\lambda_2^2} \mathbf U_2 \boxtimes \mathbf U_2 +\frac{\ln(\lambda_2^2/ \lambda_1^2)}{2(\lambda_2^2 - \lambda_1^2)} \left[ \mathbf U_1 \boxtimes \mathbf U_2 +  \mathbf U_2 \boxtimes \mathbf U_1 \right] & \text{if  } \lambda_2 \ne \lambda_1;\\
& \\
&\\
\frac{1}{2\lambda_1^2} \, \displaystyle \sum_{i=1}^{2} \sum_{j=1}^{2}\mathbf U_i \boxtimes \mathbf U_j    & \text{if  } \lambda_2 = \lambda_1; \\
& \\
\end{cases}
\end{eqnarray}
where $\mathbf U_1  = \mathbf r_1 \otimes \mathbf r_1$ and $\mathbf U_2  = \mathbf r_2 \otimes \mathbf r_2$ are second-order eigentensors of $\mathbf U$.
\item 
\begin{multline}
\mathbf{\mathcal L}^{(2)} =\frac{\partial^2 \mathbf E^{(0)} }{\partial \mathbf C \partial \mathbf C} = \begin{cases}
 -\frac{1}{2 \lambda_1^4} \bigg \lgroup \mathbf U_1 \boxtimes \mathbf U_1 \boxtimes \mathbf U_1 \bigg \rgroup -  \frac{1}{2 \lambda_2^4} \bigg \lgroup \mathbf U_2 \boxtimes \mathbf U_2 \boxtimes \mathbf U_2 \bigg \rgroup +\\  \frac{\ln(\lambda_2^2/\lambda_1^2)-(\lambda_2^2/\lambda_1^2 - 1)}{2(\lambda_2^2 -\lambda_1^2)^2}\bigg \lgroup   \mathbf U_1 \boxtimes \mathbf U_1 \boxtimes \mathbf U_2  + \mathbf U_1 \boxtimes \mathbf U_2 \boxtimes \mathbf U_1+ \mathbf U_1 \boxtimes \mathbf U_2 \boxtimes \mathbf U_2  \bigg \rgroup + \\
\frac{\ln(\lambda_1^2/\lambda_2^2)-(\lambda_1^2/\lambda_2^2 - 1)}{2(\lambda_1^2 -\lambda_2^2)^2}\bigg \lgroup   \mathbf U_2 \boxtimes \mathbf U_2 \boxtimes \mathbf U_1  + \mathbf U_2 \boxtimes \mathbf U_1 \boxtimes \mathbf U_2+ \mathbf U_2 \boxtimes \mathbf U_1 \boxtimes \mathbf U_1  \bigg \rgroup & \text{if  } \lambda_2 \ne \lambda_1; \\
 \frac{-1}{2 \lambda_1^4} \displaystyle \sum_{i=1}^{2} \sum_{j=1}^{2} \sum_{k=1}^{2} \mathbf U_i \boxtimes \mathbf U_j \boxtimes \mathbf U_k   &  \text{if  } \lambda_2 = \lambda_1.
 \end{cases}
\end{multline}
\end{enumerate}
Here `$\boxtimes$' denotes the Kronecker product $\text{Lin} \times \text{Lin} \rightarrow \mathbb L\text{in} $ that maps $n$ second-order tensors $\mathbf A$, $\mathbf B$, ..., $\mathbf C$ to a $2n$-order tensor $\mathbf A \boxtimes \mathbf B\boxtimes ... \mathbf C $. Particularly useful for this work are the following properties of the Kronecker product  \cite{kintzel2006fourthI, kintzel2006fourthII}:
\begin{equation}
(\mathbf A \boxtimes \mathbf B)\mathbf X =  \mathbf A \mathbf X \mathbf B^T,\ \forall\ \mathbf X \in \text{Lin},
\end{equation}
and 
\begin{equation}
\big(\left(\mathbf A \boxtimes \mathbf B \boxtimes \mathbf C\right)\mathbf X \big) \mathbf Y =  (\mathbf A \mathbf X \mathbf B^T \boxtimes \mathbf C ) \mathbf Y = \mathbf A \mathbf X \mathbf B^T \mathbf Y \mathbf C^T,\ \forall\  \mathbf X,\ \mathbf Y \in \text{Lin}.
\end{equation}
\\
\section{Training set for evaluation of the elastic constants}
\label{train}
The constants appearing in the expression for $\psi$ are determined via a fit to a set of  \textit{ab initio} data. This set comprises homogeneous deformations including pure equi-biaxial stretch, i.e., $\mathbf F = \mathbf U^a =J^{1/2} \mathbf I$ with the value of $J^{1/2}$ ranging from 1 to 1.2, along with an imposed isochoric shape change stretch of the form $\tilde{\mathbf U} = \lambda \mathbf r_1 \otimes \mathbf r_1 + \lambda^{-1} \mathbf r_2 \otimes \mathbf r_2$, with 
$\lambda$-values in the range  $1 \le \lambda \le \lambda_{\mathrm{crit}}(\epsilon_a)$, the upper limit being the point at which the underlying lattice becomes unstable. Only the \textit{ab initio} energies of the stable lattice configurations are considered in the dataset (see Fig.~\ref{energy}). The two principal stretch directions in the component form may be expressed as  $\mathbf r_1 = \cos \theta \,\mathbf e_1 + \sin \theta \,  \mathbf e_2 $ and  $\mathbf r_2 = -\sin \theta \, \mathbf e_1 + \cos \theta \, \mathbf e_2 $. It is noted that, owing to the $\mathcal C_{6v}$ symmetry of graphene, isochoric stretches need be sampled only over the parametric range $ 0 \le \theta \le \pi/6$. The least-squares-fit values of the material constants in the expression for $\psi$ are tabulated in Tables~(\ref{mucoeff}), (\ref{nucoeff}) and (\ref{coeffs}). \\
Our \textit{ab initio} calculations are based on first-principles density functional theory as implemented in the {\sc PWSCF} simulation package \cite{giannozzi2009quantum}. In generalized gradient approximation (GGA) calculations, the  exchange correlation energy of electrons is approximated by the generalized gradient function of Perdew, Burke, and Ernzerhof (PBE) \cite{perdew1996generalized, perdew1996rationale}, whereas in local density approximation (LDA) calculations, the exchange-correlation function of Perdew and Wang (PW) is used. The interaction between ionic cores and valence electrons is represented by an ultrasoft pseudopotential \cite{vanderbilt1990soft}. Kohn-Sham wave functions were represented using a plane-wave basis with an energy cutoff of 30 Ry and a charge density with a cutoff of 300 Ry. Integration over the irreducible Brillouin zone (BZ) for charge density and total energy was performed with a uniform $30 \times 30 \times 1$	mesh of $k$-points, and occupation numbers were smeared using the Marzari-Vanderbilt cold smearing scheme \cite{mv} with broadening of 0.03 Ry. Errors in the Cauchy stresses and total energy due to basis-set size, smearing parameter, and $k$-points are converged to less than 0.034 N/m and 0.01 Ry, respectively.\\
The phonon dispersion relations are computed via linear response calculations as implemented in density functional perturbation theory \cite{DFPT, DFPT2}. The dynamical matrix  is calculated on  an $8 \times 8 \times 1 $ uniform grid of 
$\mathbf q$-points in the irreducible BZ using a 30 $\times$ 30 $\times$ 1 uniform grid of $k$-points.  The dynamical matrix is fast-Fourier-transformed to calculate the interatomic force constants (IFC), corrected for acoustic sum rule to ensure that $\omega(\mathbf q=\mathbf 0)=0$ for all the acoustic branches. The IFC's are then used to interpolate the phonon frequencies over a dense set of $\mathbf q$-points along the high-symmetry directions in the irreducible BZ at different deformed states. All the calculations in this work are performed on a two-atom primitive unit cell of graphene shown in  Fig.~(\ref{lattice}). The two lattice vectors defining the undeformed unit cell are $\mathbf a_1 = a_0 \bigg \lgroup \frac{\sqrt{3}}{2} \mathbf e_1 + \frac{1}{2} \mathbf e_2\bigg \rgroup$, and $\mathbf a_2 = a_0  \bigg \lgroup \frac{\sqrt{3}}{2} \mathbf e_1 - \frac{1}{2} \mathbf e_2\bigg \rgroup$, where $a_0$ is the lattice constant. The LDA and GGA values of the undeformed lattice constants are  $2.44 \mathring{A}$ and $2.465\mathring A$, respectively, both of which are very close to the 
experimentally-reported value\cite{baskin1955lattice} of 2.457 $\mathring{A}$. \\
\begin{minipage}{\linewidth}
\centering
\includegraphics[scale=0.6]{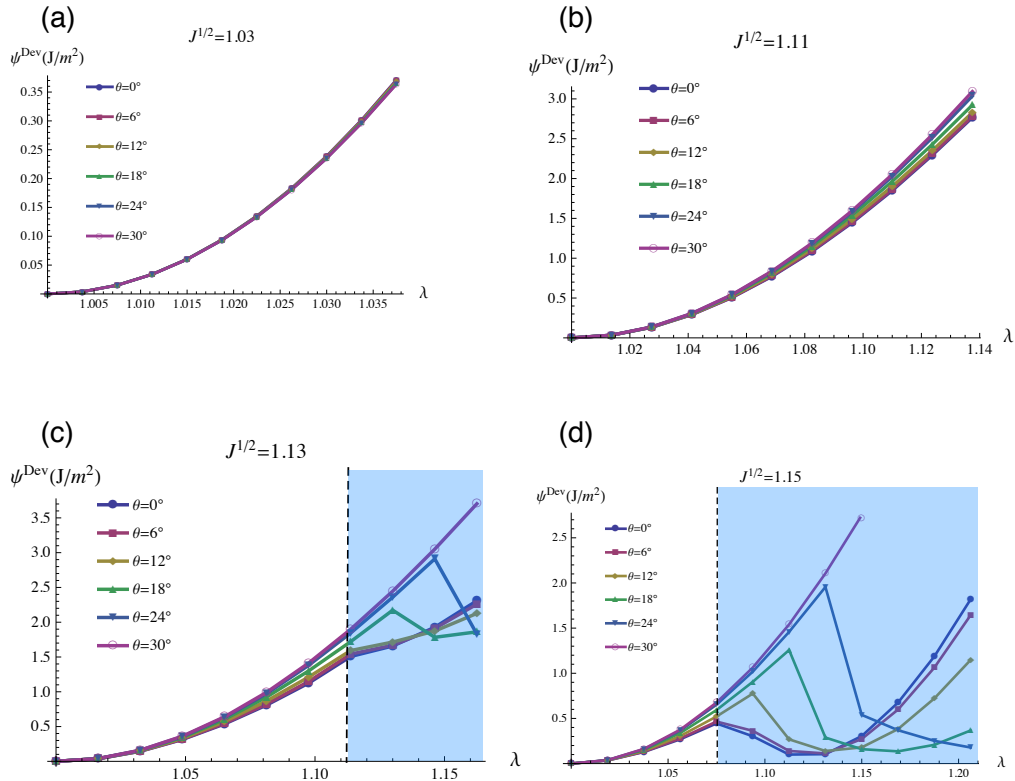}
\captionof{figure}{Plots of the \textit{ab initio} deviatoric energy $\psi^{\mathrm{Dev}}$ as a function of shape-changing stretch $\lambda$ at various values of biaxial stretch $J^{1/2}$: (a) \& (b) --- Graphene's mechanical response remains essentially isotropic at small to moderate strains, with anisotropy appearing only at large deviatoric stretch. (c) \& (d) --- For larger values of biaxial strain,  sudden drops in $\psi^{\mathrm{Dev}}$, indicative of material instability, are noted at finite $\lambda$-values. Data from the unstable regime (shown as blue-shaded) is not included in the training set used for deriving the constitutive model.}
\label{energy}
\end{minipage} \\  \\
\section{Validation of the constitutive model}
\label{valid}
As detailed in the following sections, we validate our continuum model for a number of homogeneous deformations by comparison with small-strain elastic constants inferred from experiments, comparison of stress-strain curves with independent \textit{ab initio} calculations, and comparisons of predictions of elastic stability limits to independent phonon calculations. 
\subsection{Comparison of small-strain elastic constants with experiments}
We compare the in-plane elastic constants --- recovered from the constitutive model  in the limit of infinitesimal strain --- with the measured values of Lee, \textit{et al.} \cite{lee2008measurement} .  The predicted values for the in-plane Young's modulus $Y^{(0)}$,   shear modulus $\mu^{(0)}$,   bulk modulus $\kappa^{(0)}$, and the Poisson's ratio $\nu^{(0)}$ are all in good accord with the experimentally-reported values (see Table~[\ref{Exp}]).\\
\begin{minipage}{\linewidth}
\centering
\begin{tabular}{c c c c c}
\toprule[1.5pt] 
 & $Y^{(0)} $(N/m) & $\nu^{(0)}$& $\kappa^{(0)}$(N/m) & $\mu^{(0)}$(N/m) \\
\hline 
\hline
Exp. & $342\pm 30$ &  0.165& $205\pm18 $ & $147 \pm 12$ \\
\hline
DFT (GGA) & $ 349$ &  0.203 & $219$ & $145$ \\
\hline
DFT (LDA) & $354$ &  0.203 & $222$ & $147$ \\
\bottomrule[1.65pt]
\end{tabular}
\captionof{table}{ The in-plane elastic constants --- recovered from our constitutive model  in the limit of infinitesimal strain --- compared with values reported by Lee, \textit{et al.} \cite{lee2008measurement}, based on their experimental results.}
\label{Exp}
\end{minipage}

\subsection{Prediction of elastic stability limits}
 Material elastic stability requires that the speed of propagation of the acceleration waves in a solid in every direction should be non-negative \cite{hill1977principles}. This condition is satisfied when the acoustic tensor is positive-definite, which is  equivalent to the condition that the continuum elasto-dynamic equations are hyperbolic. \\

The acoustic tensor-based prediction of elastic instability involves detecting a deformed state at which $\mathbb A$ loses positive-definiteness \cite{Born1,Born2}  (or, equivalently, when the slope of an acoustic phonon branch in the long wavelength limit vanishes, i.e.,  $ d\omega/d k \rvert_{\mathbf k \rightarrow 0} =0$). In addition to elastic instabilities, at very large strains there may also be unstable optical phonon modes in the lattice,  causing abrupt rearrangement of the atoms within the basis set of the unit cell. Such long-wavelength optical instabilities ---  which can not be detected by the acoustic tensor analysis --- are accompanied by abrupt changes in energy/stress of the deformed state; e.g., as shown in the excluded regions of Fig.~[\ref{energy}].  Employing the continuum expression for $\mathbb A$ (Eq.~(\ref{acoustictensor})), we can monitor its positive-definiteness  at all points along a deformation path, and thus can precisely locate the initial loss of elastic stability in that mode of deformation. In the following subsections, we predict  elastic stability limits for some important homogeneous deformation modes, and compare these with  predictions obtained from  independent $\textit{ab initio}$ linear response phonon calculations. \\
\subsubsection{Pure biaxial stretch}
For this deformation state, Cartesian components of $\mathbb A$ obtained from the present constitutive model are
\begin{equation}
\mathbb A_{ijkl} =  \mathcal E \alpha^2 (1-\epsilon_a \alpha) e^{-\epsilon_a(1+\alpha)} \delta_{ij}\delta_{kl} +\mu(\epsilon_a) e^{-\epsilon_a} (\delta_{ik}\delta_{jl} -\delta_{kl}\delta_{ij} +\delta_{il}\delta_{jk}  )-\mathcal E \alpha^2 \epsilon_a e^{-\epsilon_a(1+\alpha)} \delta_{jk}\delta_{il}.
 \label{acoustic2}
\end{equation}
The condition of strong ellipticity requires 
\begin{equation}
(\mathbf m \otimes \mathbf n ): \mathbb A : (\mathbf m \otimes \mathbf n) =  \mathbb A_{ijkl} m_i m_k n_j n_l  > 0,
\label{strongellipticity}
\end{equation}
for all arbitrary unit vectors $\mathbf m$ and $\mathbf n$. If, for any two unit vectors, Eq.~(\ref{strongellipticity}) does not hold, then the deformed crystal is said to have lost strong ellipticity.
Substituting for $\mathbb A$ from Eq.~(\ref{acoustic2}), the strong ellipticity condition becomes
\begin{equation}
(\mathbf m \otimes \mathbf n ): \mathbb A : (\mathbf m \otimes \mathbf n)= e^{-\epsilon_a(1+\alpha)} \mathcal E \alpha^2 \big(1- \epsilon_a(1+\alpha)\big) \, m_i n_i m_j n_j +  e^{-\epsilon_a}\mu (\epsilon_a)\,  m_i m_i n_i n_j > 0.
 \label{SE_biax}
\end{equation}
Let the unit vectors $\mathbf m$ and $\mathbf n$ be represented by
\begin{equation}
\mathbf m = ( \cos \phi, \sin \phi );\ \mathbf n = ( \cos \varphi, \sin \varphi );
\end{equation}
using this parametrization in Eq.~\ref{SE_biax} provides
\begin{equation}
(\mathbf m \otimes \mathbf n ): \mathbb A : (\mathbf m \otimes \mathbf n)=e^{-\epsilon_a(1+\alpha)} \mathcal E \alpha^2 \big(1- \epsilon_a(1+ \alpha)\big)\, \cos^2 (\phi -\varphi)+ 
 e^{-\epsilon_a}\,\mu(\epsilon_a) >0.
\end{equation}
When $\epsilon_a < 1/(1+\alpha) = \epsilon_a|_{\kappa=0}$, the minimum value of the term containing the expression $\cos^2 (\phi -\varphi)$ is zero, occurring when $\mathbf m . \mathbf n = 0$; therefore 
 \begin{equation}
\left[(\mathbf m \otimes \mathbf n ): \mathbb A : (\mathbf m \otimes \mathbf n)\right ]_{\mathrm{min}}=   e^{-\epsilon_a}\mu(\epsilon_a).
\end{equation}
Thus, the loss of strong ellipticity under equi-biaxial stretch first occurs when (also see Table~(\ref{mucoeff}))
\begin{equation}
\mu(\epsilon_a) = \mu_0 - \mu_1 e^{\beta \epsilon_a} =0.
\end{equation}
This condition occurs at a critical value of $\epsilon_a$ given by
\begin{equation}
\epsilon_a\rvert_{\mu=0} = \frac{1}{\beta } \ln \left( \frac{\mu_0}{\mu_1} \right).
\end{equation}
 Based on values of the fitted constants, $\epsilon_a\rvert_{\mu=0} < \epsilon_a\rvert_{\kappa=0}$, as assumed.  Since the associated directions $\mathbf m$ and $\mathbf n$ are perpendicular to each other, the dynamic instability is of transverse acoustic nature. To independently confirm this prediction, we carry out  linear-response-based phonon calculations for equi-biaxially strained graphene. The phonon dispersion shows that the strong ellipticity condition ceases to hold at $\epsilon_a  =\ln J=0.354$, at which point a long-wavelength transverse acoustic instability appears, due to the vanishing of $\mu$ (Fig.(\ref{acoustic})). Based on our constitutive modeling and on independent phonon calculations, we see the loss of elastic stability under equi-biaxial loading occurring prior to the zero tangent modulus condition. \\
 
\begin{minipage}{\linewidth}
\centering
\includegraphics[scale=0.7]{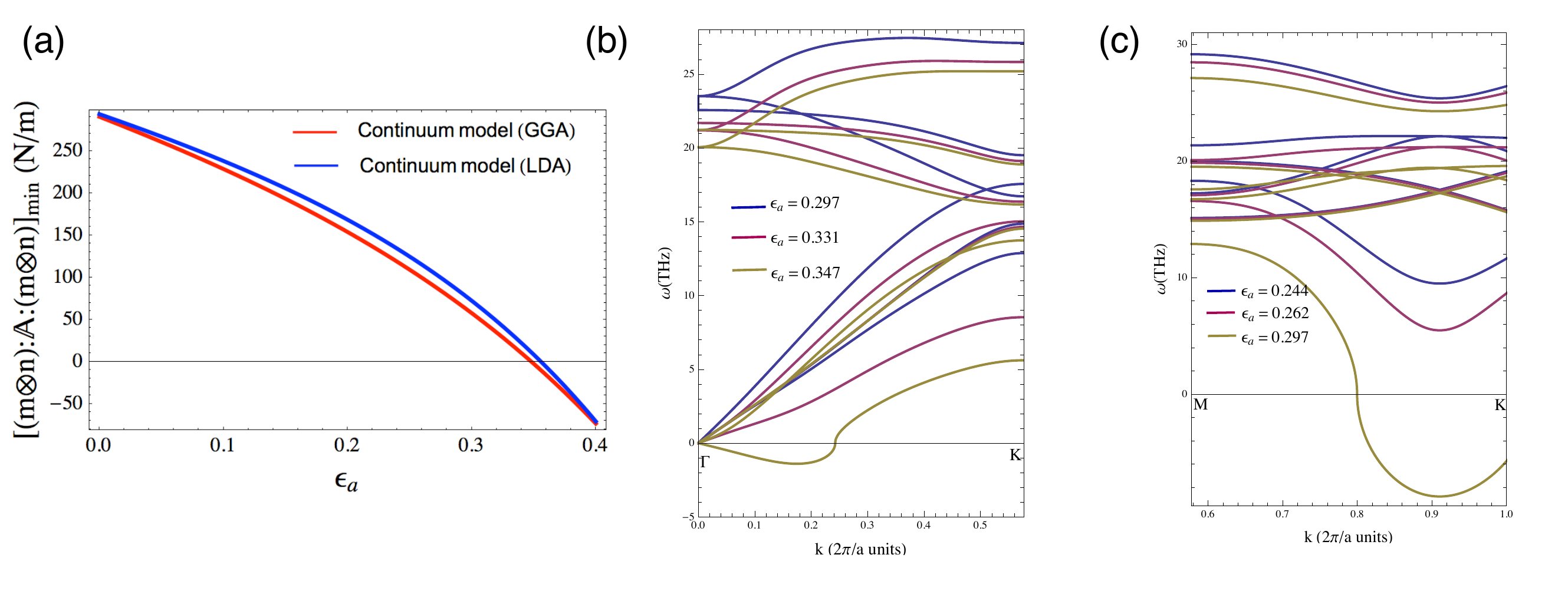}
\captionof{figure}{ (a) The instability due to loss of positive-definiteness of $\mathbb A$ in equi-biaxial stretch ($\mathbf F = J^{1/2} \mathbf I$) corresponds to the vanishing of $\mu(\epsilon_a)$ that occurs at the critical equi-biaxial areal strain of $\epsilon_a=0.35$(GGA) and $0.36$(LDA). (b) LDA Phonon dispersion along the $\Gamma- K$ direction shows a long-wavelength transverse acoustic phonon at $\Gamma$ going unstable near $\epsilon_a = 0.34$; softening of this branch is associated with the vanishing shear modulus $\mu(\epsilon_a)$. (c) Phonon dispersions along the $M- K$ direction show a soft-phonon mode appearing just before  $\epsilon_a = 0.297$, well before the acoustic instability. }
\label{acoustic}
\end{minipage}\\
\subsubsection{Uniaxial strain in armchair and zigzag directions}
Uniaxial stretching along both armchair and zigzag directions preserves the reflection symmetries w.r.t  the $\hat{\mathbf {x}}$ and $\hat{\mathbf{y}}$-axes. Therefore, for such deformations, the $ \mathbb A_{1112}$, $\mathbb A_{1121}$, $\mathbb A_{1211}$, $\mathbb A_{2111}$, $ \mathbb A_{2212}$, $\mathbb A_{2221}$, $\mathbb A_{1222}$, and $\mathbb A_{2122}$ components of the acoustic tensor are all zero. The resultant acoustic tensor can be represented by a block-diagonal $4 \times 4$ matrix of the form:
\begin{equation}
\mathbb A = \begin{bmatrix}
\mathbb A_{1111} & \mathbb A_{1122} & 0 & 0 \\
\mathbb A_{1122} & \mathbb A_{2222} & 0 & 0 \\
0 & 0 & \mathbb A_{1212} & \mathbb A_{1221} \\
0 & 0 & \mathbb A_{1221} & \mathbb A_{2121} \\
\end{bmatrix}.
\label{acoustict}
\end{equation}
 The eigenvalues of this matrix are given by
\begin{equation}
\Lambda_{1} = \frac{1}{2} \left[ \mathbb A_{1111}+ \mathbb A_{2222} + \left( (\mathbb A_{1111} - \mathbb A_{2222})^2 + 4 \mathbb A_{1122}^2 \right)^{1/2} \right],
\end{equation}
\begin{equation}
\Lambda_{2} = \frac{1}{2} \left[ \mathbb A_{1111}+ \mathbb A_{2222} - \left( (\mathbb A_{1111} - \mathbb A_{2222})^2 + 4 \mathbb A_{1122}^2 \right)^{1/2} \right],
\end{equation}
\begin{equation}
\Lambda_{3} = \frac{1}{2} \left[ \mathbb A_{1212}+ \mathbb A_{2121} - \left( (\mathbb A_{1212} - \mathbb A_{2121})^2 + 4 \mathbb A_{1221}^2 \right)^{1/2} \right],
\end{equation}
and
\begin{equation}
\Lambda_{4} = \frac{1}{2} \left[ \mathbb A_{1212}+ \mathbb A_{2121} + \left( (\mathbb A_{1212} - \mathbb A_{2121})^2 + 4 \mathbb A_{1221}^2 \right)^{1/2} \right].
\end{equation}
At the first loss of strong ellipticity, one or more eigenvalues of the acoustic tensor matrix become zero at a certain critical value of $\lambda$. In order to determine the critical stretch at which $\mathbb A$ loses  positive-definiteness, we track the variation of the eigenvalues $\Lambda_1$, $\Lambda_2$, $\Lambda_3$ and $\Lambda_4$ with increasing uniaxial stretch. \\
First we consider  uniaxial stretching along the zigzag direction,  oriented along the Cartesian $\mathbf e_1$-axis, with $\mathbf U =  \lambda_s \mathbf e_1 \otimes \mathbf e_1+ 1\, \mathbf e_2 \otimes \mathbf e_2 $. For this case, the loss of strong ellipticity occurs at  a critical stretch, $\lambda_s \approx 1.18$ for GGA, and $\approx 1.19$ for LDA, when the $\Lambda_2$-eigenvalue of the acoustic tensor vanishes, as shown in Fig.~(\ref{zigzaguniaxialstrain}-a). The associated eigenvector corresponds to $\mathbf m = \mathbf n = \mathbf e_1$, (see Eq.~(\ref{strongellipticity})), so the unstable longitudinal mode coincides with the maximum principal eigenvector of $\mathbf U$. From independent phonon dispersion calculations shown in Fig.~(\ref{zigzaguniaxialstrain}-b), we confirm the occurrence of a longitudinal acoustic instability in this direction,  at $\lambda_s \approx 1.188$.\\ 

\begin{minipage}{\linewidth}
\centering
\includegraphics[scale=0.6]{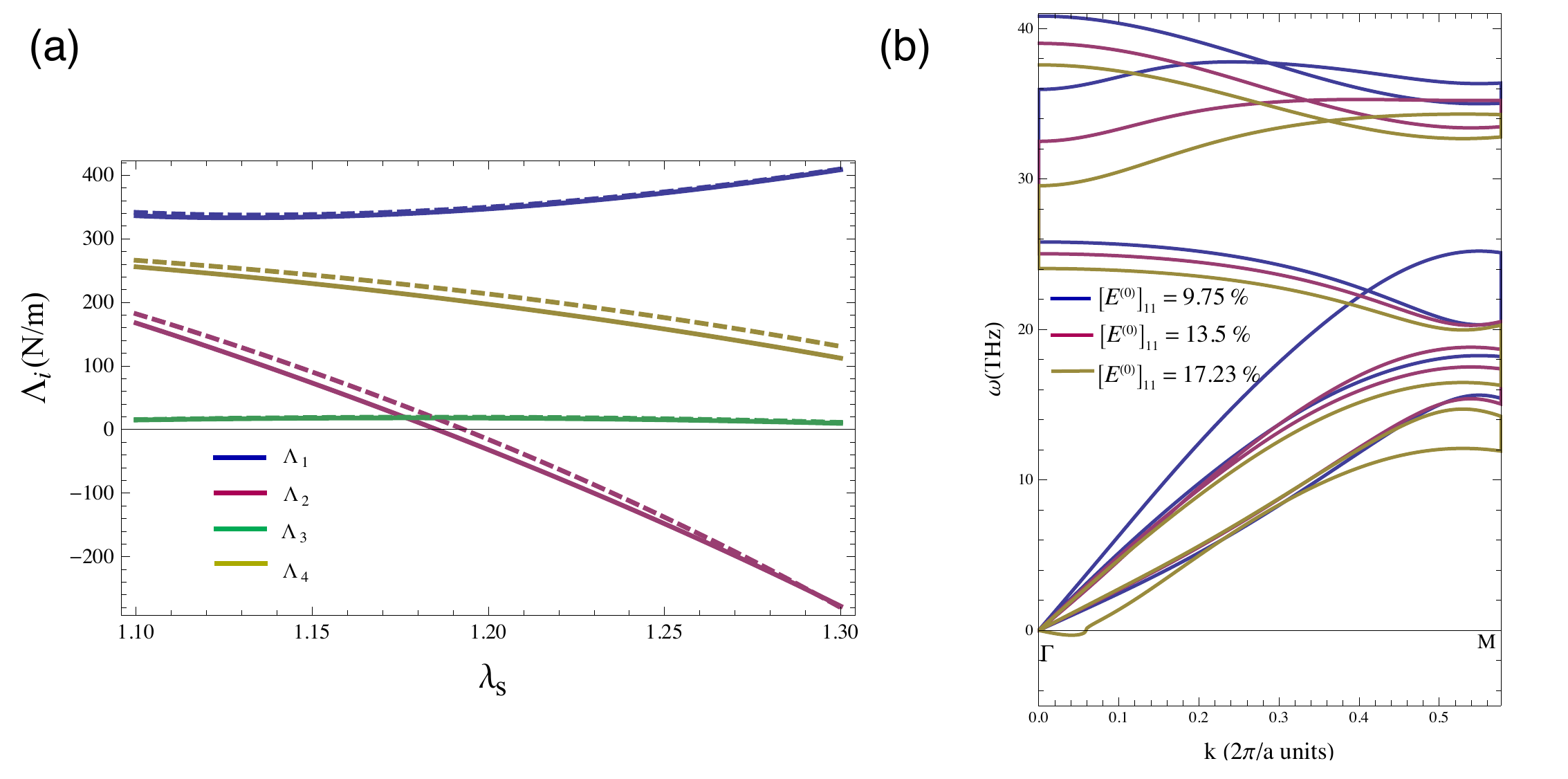}
\captionof{figure}{ (a) Eigenvalues of the acoustic tensor $\mathbb A$ as a function of stretch $\lambda_s$ in the zigzag direction. The eigenvalue $\Lambda_2$ vanishes at a critical stretch,  $\lambda_s \approx 1.18$ for GGA, and $\approx 1.19$ for LDA, indicating an acoustic instability in the material. The associated eigenvector of  the unstable mode occurs in the $\mathbf e_1$-direction with polarization along $\mathbf e_1$-direction, implying  longitudinal instability mode. The solid/dashed lines are from constitutive fits to GGA/LDA results. (b) LDA phonon dispersion curves along the $\Gamma - M$ direction at increasing values of uniaxial strain. A long-wavelength longitudinal instability emerges at a uniaxial logarithmic strain of $17.23\%$, 
corresponding to $\lambda_s =1.188$.}
\label{zigzaguniaxialstrain}
\end{minipage}
\\ \\
Now consider uniaxial uniaxial stretching along the armchair direction, which is taken to be oriented along the Cartesian $\mathbf e_2$-axis, with $\mathbf U = 1 \,\mathbf e_1 \otimes \mathbf e_1+\lambda_a\, \mathbf e_2 \otimes \mathbf e_2 $. In this case, $\mathbb A$ loses positive-definiteness at  $\lambda_a \approx 1.23$ (GGA ) and $\approx 1.24$ (LDA), when the $\Lambda_2$-eigenvalue of the acoustic tensor goes to zero, as shown in Fig.~(\ref{armchairuniaxialstrain}-a). The associated eigenvector in this case corresponds to $\mathbf m = \mathbf n = \mathbf e_2$,
so in this case as well, a longitudinal elastic instability occurs in the direction of maximum principal stretch. This result is again confirmed by independent LDA phonon calculations shown in Fig.~(\ref{armchairuniaxialstrain}-b), which indicate a LA branch in the $\Gamma-K$ direction with a vanishing slope at $\Gamma$ emerging when $\lambda_a = 1.252$.
\\ \\
  \begin{minipage}{\linewidth}
\centering
\includegraphics[scale=0.6]{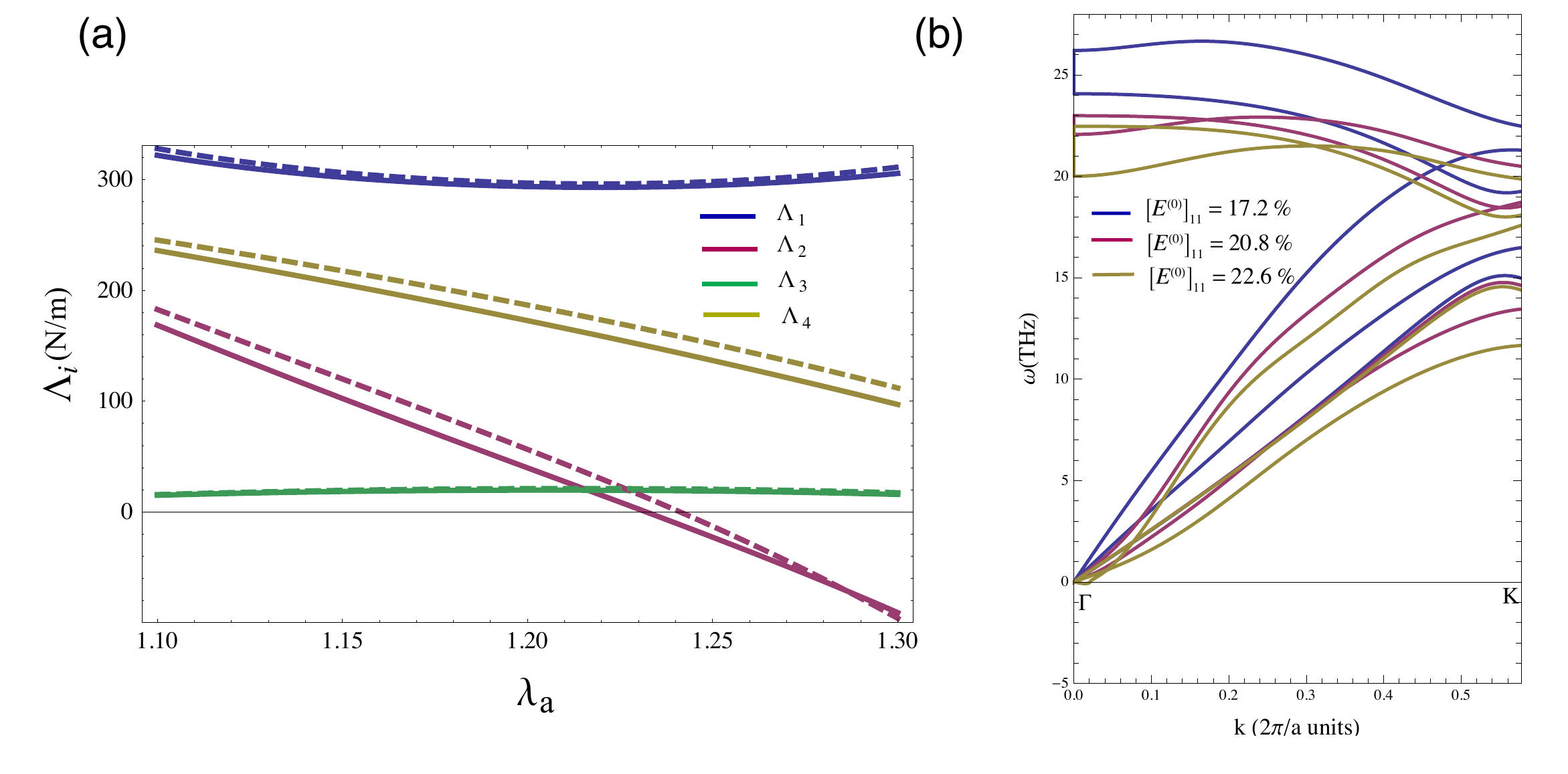}
\captionof{figure}{(a) Eigenvalues of the acoustic tensor $\mathbb A$ as a function of stretch $\lambda_a$ along the armchair direction parallel to $\mathbf e_2$. The eigenvalue $\Lambda_2$ vanishes at a critical stretch $\lambda_a \approx 1.23$  (GGA) and $ \approx 1.24$ for (LDA), indicating an acoustic instability in the material. The associated eigenvector shows that this instability occurs in the $\mathbf e_2$-direction, and the polarization of the unstable mode is also along the $\mathbf e_2$-direction, implying that the instability is of longitudinal nature. The solid/dashed lines are calculated from constitutive fits to GGA/LDA results. (b) LDA Phonon dispersions along $\Gamma - K$ direction at increasing uniaxial strain. A long wavelength longitudinal instability emerges at a uniaxial logarithmic strain of $22.5\%$, corresponding to $\lambda_a=1.252$. }
\label{armchairuniaxialstrain}
\end{minipage} \\ \\
\subsubsection{Uniaxial stress in armchair and zigzag directions}
In cases of uniaxial stress along the armchair and zigzag directions, the resultant acoustic tensor has the same form as shown in Eq.(\ref{acoustict}), and the corresponding eigenvalues are given by Eq.~(68-71).
Uniaxial tensile stress along the zigzag direction corresponds to the right stretch tensor given by  $\mathbf U =  \lambda_s \, \mathbf e_1 \otimes \mathbf e_1+ f(\lambda_s) \, \mathbf e_2 \otimes \mathbf e_2, $ where the transverse stretch $f(\lambda_s)\le1$, and its value, for a given $\lambda_s$, is determined by setting $\sigma_{22} = \mathbf{e}_2.\pmb{\sigma}.\mathbf{e}_2 =0$.
For uniaxial stress along the zigzag direction, the loss of strong ellipticity occurs at $\lambda_s= 1.19$, when the $\Lambda_2$-eigenvalue of $\mathbb A$ becomes zero (see Fig.~(\ref{zigzaguniaxialstress})). The associated eigenvector again corresponds to $\mathbf m =\mathbf n =\mathbf e_1$, so that the unstable longitudinal mode  parallels the principal stretching eigenvector.  Independent phonon calculations confirm the occurrence of an acoustic instability in the longitudinal branch of the phonon dispersion at $\lambda_s \approx 1.20$, as was also indicated by  phonon calculations of Liu, \textit{et al.} \cite{liu2007ab}.\\
For uniaxial tensile stress along the armchair direction,  $\mathbf U =  g(\lambda_a)\, \mathbf e_1 \otimes \mathbf e_1+\lambda_a \, \mathbf e_2 \otimes \mathbf e_2, $ where the transverse stretch $g(\lambda_a)\le1$,  its value, for a given $\lambda_a$, being determined by setting $\sigma_{11}=0$. In this case, the acoustic tensor analysis shows that the lattice instability takes place at $\lambda_a = 1.24$, as shown in Fig.~(\ref{armchairuniaxialstress}-a). This instability is also  longitudinal, occurring in the $\mathbf e_2$-direction, and is also in good agreement with the phonon calculations of Fig.~(\ref{armchairuniaxialstress}-b). \\

\begin{minipage}{\linewidth}
\centering
\includegraphics[scale=0.6]{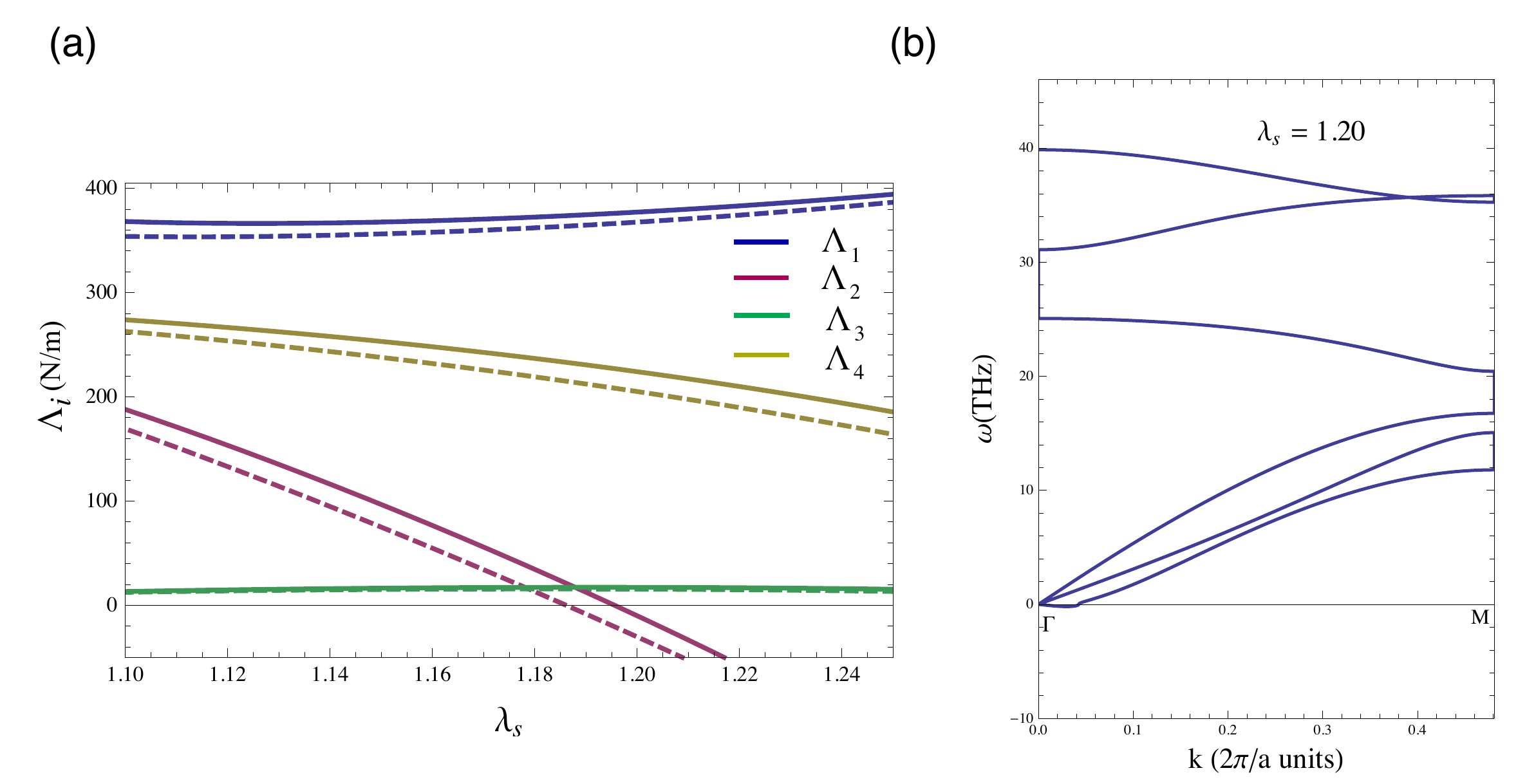}
\captionof{figure}{(a) Eigenvalues  of the acoustic tensor $\mathbb A$ as a function of stretch $\lambda_s$ along the zigzag direction. The eigenvalue $\Lambda_2$ vanishes at a critical value of $\lambda_s \approx 1.186$ for GGA and $\lambda_s \approx 1.192$ for LDA,  indicating an acoustic instability. The associated eigenvector indicates that this instability occurs in the $\mathbf e_2$-direction, and the polarization of the unstable mode is also along the $\mathbf e_2$-direction, implying that the instability is of longitudinal nature. The solid/dashed lines are from constitutive fits to GGA/LDA results.  (b) LDA Phonon dispersions along the $\Gamma - M$ direction at increasing uniaxial stress in the zigzag direction. A long-wavelength instability with longitudinal polarization emerges at $\lambda_s=1.20$.}
\label{zigzaguniaxialstress}
\end{minipage}
\\ 

\begin{minipage}{\linewidth}
\centering
\includegraphics[scale=0.6]{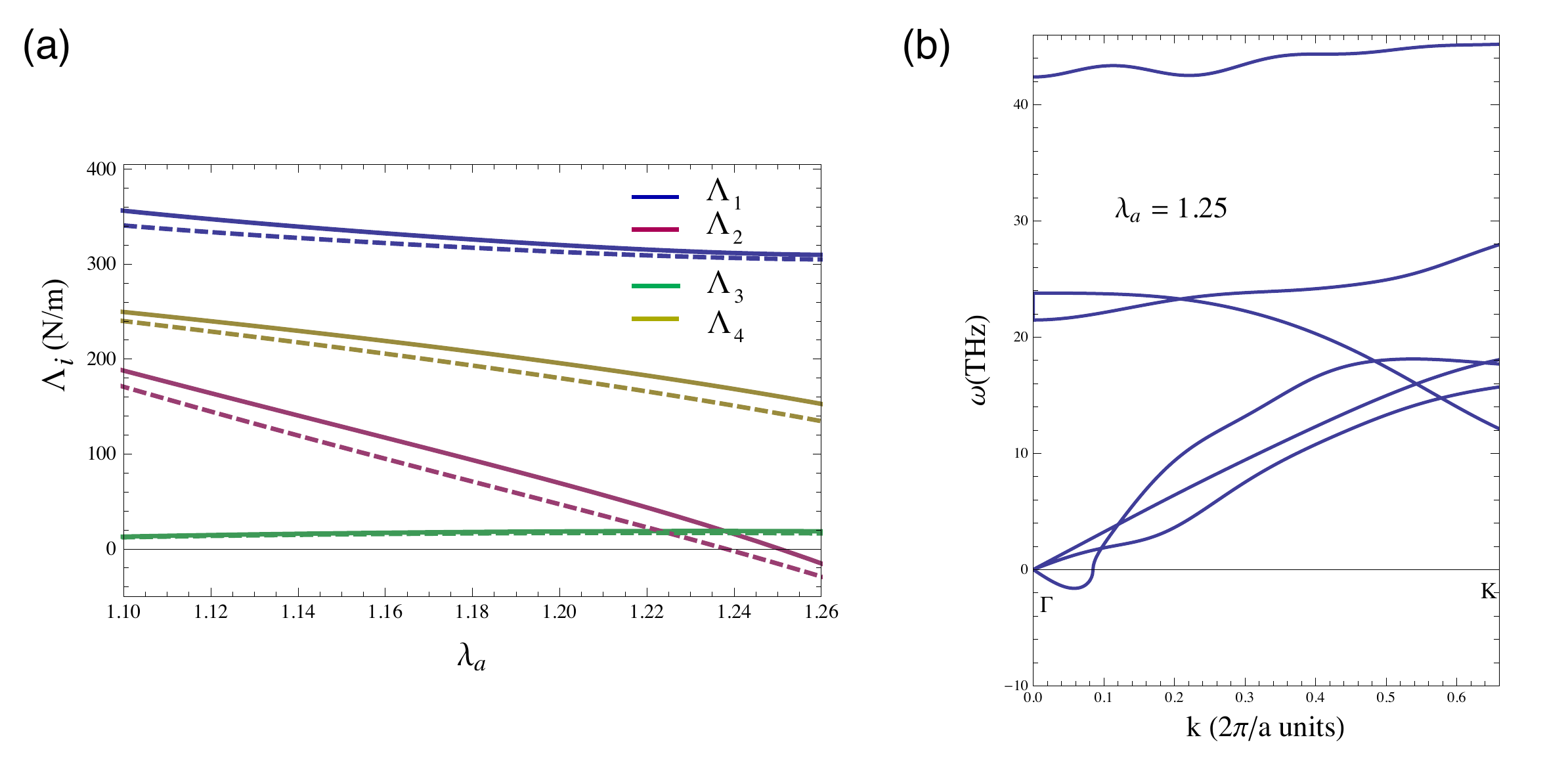}
\captionof{figure}{(a) Eigenvalues of the acoustic tensor $\mathbb A$ as a function of the uniaxial stretch $\lambda_a$ along the armchair direction. The eigenvalue $\Lambda_2$ vanishes at a critical value of  $\lambda_a \approx1.238$ for GGA and $\approx 1.25$ for LDA . This indicates an acoustic instability in the material. From the associated eigenvector, we infer that this instability occurs in the $\mathbf e_2$-direction, and the polarization of the unstable mode is also along the $\mathbf e_2$-direction, implying that the instability is of longitudinal nature. The solid/dashed lines are from constitutive fits to GGA/LDA results. (b) LDA Phonon dispersions along the $\Gamma - K$ direction at increasing uniaxial stress along the armchair direction. A long wavelength instability with longitudinal polarization emerges at $\lambda_a=1.25$.  }
\label{armchairuniaxialstress}
\end{minipage}
\\ \\
The predicted elastic stability limits  for various deformation modes considered in this work are summarized in Table~[\ref{coeffs}]. Particularly, we note that the stretch limits for elastic stability under both uniaxial stress and stretch, in both the zigzag and armchair directions, are very close; this must be attributed to the fact that graphene has a small Poisson ratio, and the ratio of the magnitude of transverse to axial log strains under uniaxial stress further decreases as the lattice is deformed
 (see Fig.~(\ref{poissonlike})). 
\begin{table}[htbp!]
\begin{tabular}{l l  l  }
\toprule[1.5pt]
Deformation mode & Acoustic tensor analysis & Phonon analysis (LDA) \\
\hline 
\hline
\\
Equi-biaxial stretch & \parbox{2in}{Elastic shear instability at \\ $J^{1/2} = 1.19$}  & \parbox{2in}{Short-wavelength transverse instability at $J^{1/2} = 1.145$, and \\  Transverse acoustic instability at $J^{1/2} = 1.18$} \\ \\
\hline
\\ 
Uniaxial stretch (zigzag) & \parbox{2in}{Elastic longitudinal instability at $\lambda_s =1.188$(GGA)\\ ($\lambda_s = 1.194$(LDA)) } & \parbox{2in}{ Longitudinal acoustic instability at $\lambda = 1.188 $ } \\ \\  
\hline
\\
Uniaxial stretch (armchair) &  \parbox{2in}{Elastic longitudinal instability at $\lambda_a = 1.23$ (GGA)\\ ($\lambda_a = 1.24$(LDA))}  &  \parbox{2in}{ Longitudinal acoustic instability at $\lambda = 1.25 $ }  \\  \\
\hline
\\
Uniaxial stress (zigzag) &  \parbox{2in}{Elastic longitudinal instability at $\lambda_s = 1.186$(GGA)\\($\lambda_s = 1.192$(LDA)) }  &  \parbox{2in}{ Longitudinal acoustic instability at $\lambda = 1.20 $ }  \\  \\
\hline
\\
Uniaxial stress (armchair) &  \parbox{2in}{Elastic longitudinal instability at $\lambda_a = 1.238$(GGA) \\ ($\lambda_a = 1.25$(LDA))}  &  \parbox{2in}{ Longitudinal acoustic instability at $\lambda = 1.25 $ }  \\  \\
\bottomrule[1.5pt]
\end{tabular}
\caption{Summary of instability analyses based on acoustic tensor analysis of the constitutive models and the corresponding phonon calculations for various homogeneous deformation modes considered.}
\label{coeffs}
\end{table}

\section{Stress-strain response curves}
\label{strstr}
In the present approach the coefficients of the continuum model are determined by a least-squares fit to the strain energies alone, and the stresses were not used in the fitting, so the agreement of predicted stress values  with those determined directly from \textit{ab initio} calculations remains to be examined. In the following subsections we compare the continuum model's predictions of  stress-strain response in a few important modes of large homogeneous deformation modes with corresponding \textit{ab initio}-calculated values. For this purpose,  we use the continuum model to obtain the components of stress as functions of strain for a set of deformations  including equi-biaxial tension, uniaxial stretching in the armchair, and the zigzag directions, uniaxial tension in the armchair and the zigzag directions, and compare the predicted stresses with the values directly calculated from first-principles calculations. In obtaining the stress-strain curves, the elastic and soft-mode instabilities --- discussed in previous sections --- have been suppressed in both \textit{ab initio} and in continuum calculations.
\subsection{Pure biaxial stretch}
The Cauchy stress as a function of strain for pure equi-biaxial deformation as obtained from the continuum model is shown in Fig.[\ref{sigma}-a].  The Cauchy biaxial stress as a function of the areal strain, as predicted by the UBER-based model, compares with the \textit{ab initio} values well. Our model predicts that the Cauchy stress reaches its maximum value at nearly $42\%$ areal strain, beyond which graphene becomes elastically unstable with respect to to pure areal deformation. In Fig.[\ref{sigma}-b], we have shown the softening of the dilatant tangent modulus, defined in Eq.~(\ref{bm}), with areal strain $\epsilon_a$. The areal tangent modulus vanishes when the Cauchy stress approaches its maximum value. \\

\begin{minipage}{\linewidth}
\centering
\includegraphics[scale=0.6]{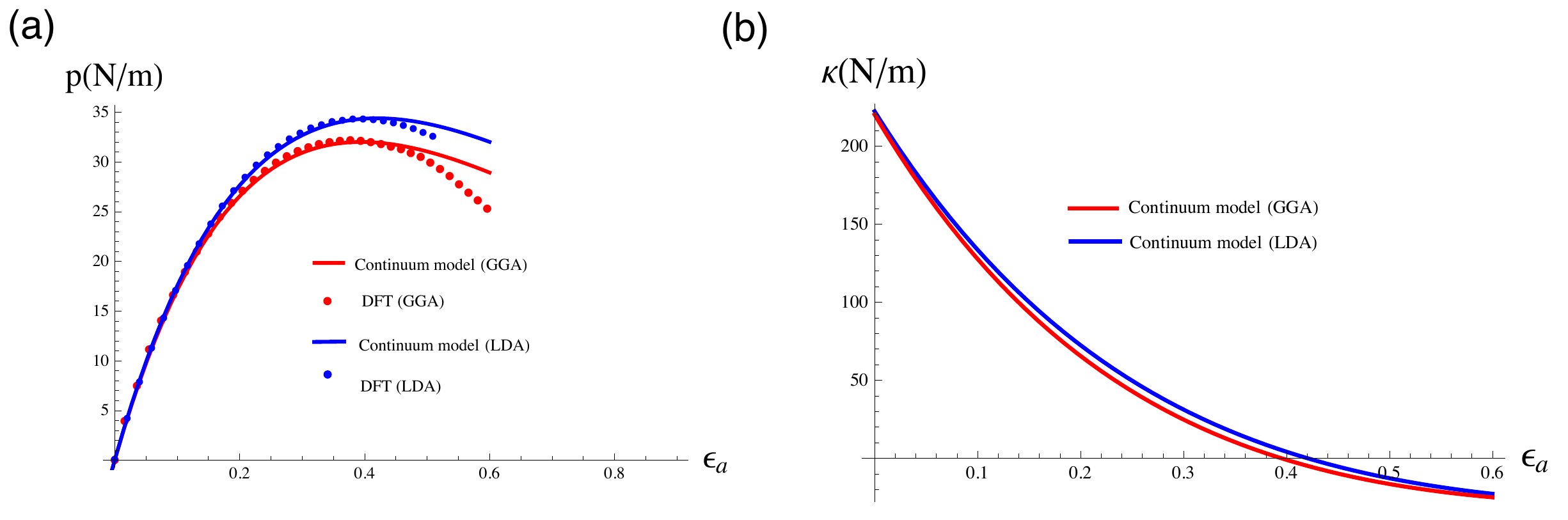}
\captionof {figure}{ (a) Variation of Cauchy mean normal stress with areal strain as predicted by the UBER-based continuum model, along with corresponding \textit{ab initio} values. (b) Variation of the dilatational tangent modulus $\kappa$ with areal strain $\epsilon_a = \ln J$, as predicted by the UBER-based continuum model.}
\label{sigma}
\end{minipage}
\\  

\subsection{Uniaxial stretching along  the armchair and the zigzag directions}
The Cauchy stress as a function of strain obtained from the continuum model for the case of uniaxial stretching along the armchair and the zigzag directions are shown in Fig. [\ref{armchairuniaxialstrainstressstrain}-a] and Fig. [\ref{armchairuniaxialstrainstressstrain}-b], respectively.\\

\begin{minipage}{\linewidth}
\centering
\includegraphics[scale=0.6]{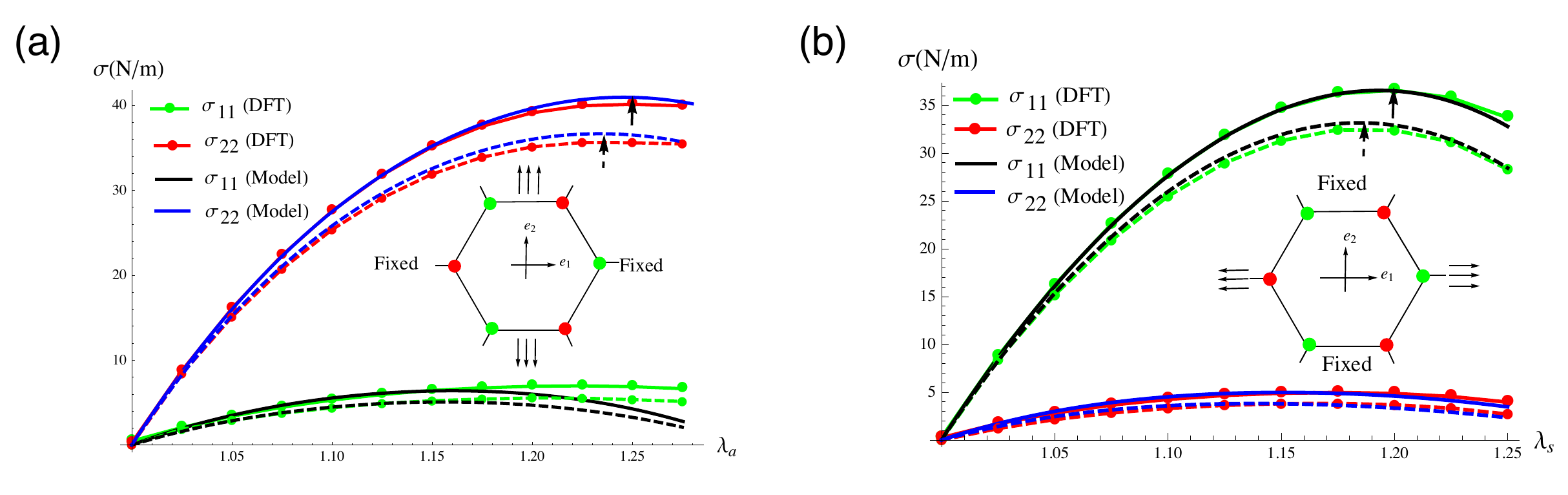}
\captionof{figure}{Stress-strain curve obtained from the continuum model for uniaxial strain along (a) armchair direction and (b) zigzag direction.  For comparison,  we have also shown the \textit{ab initio} stress values. The arrow shows the location of maximum stress. The solid lines are LDA results, whereas the dashed lines are GGA counterparts.}
\label{armchairuniaxialstrainstressstrain}
\end{minipage}
\\
\subsection{Uniaxial tension along  the armchair and the zigzag directions}
Cauchy stress components as  functions of strain obtained from the continuum model for the case of uniaxial tensile stress along the armchair and the zigzag directions are shown in Fig. [\ref{missinglabel}-a] and Fig. [\ref{missinglabel}-b], respectively.
\\
\begin{minipage}{\linewidth}
\centering
\includegraphics[scale=0.6]{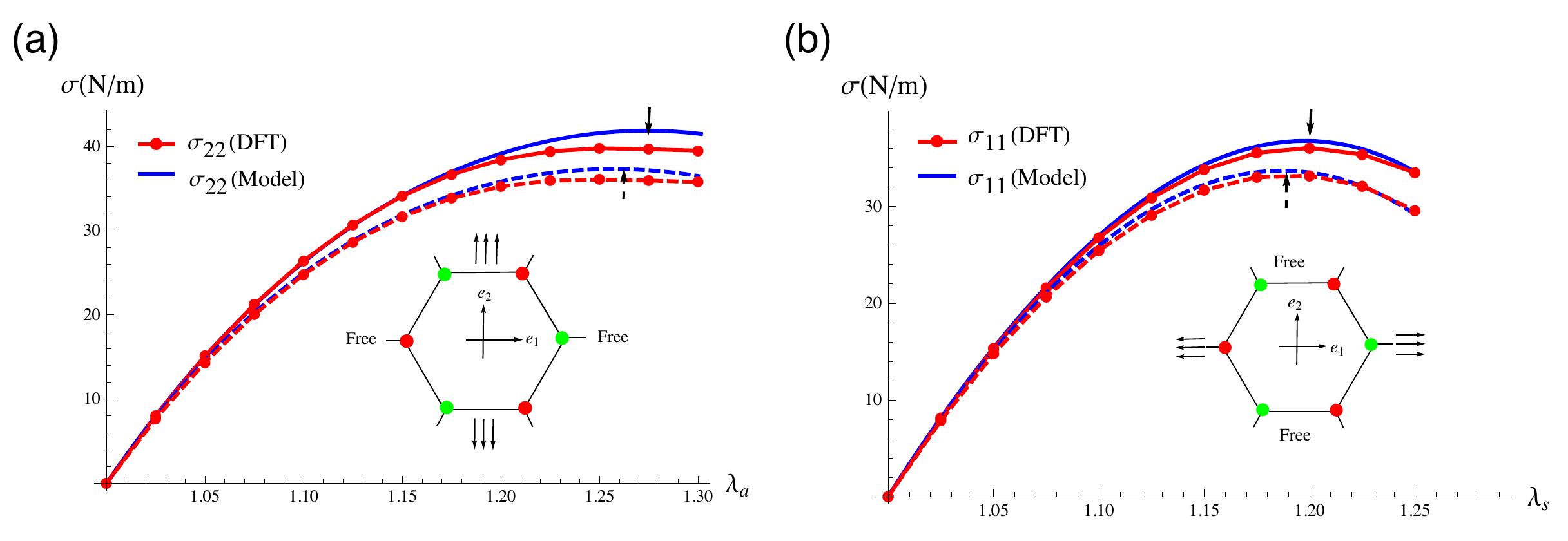}
\captionof{figure}{Validation of the continuum model for uniaxial stress along  (a) the armchair direction and (b) the zigzag direction.  The continuum model (solid line) is in good agreement with the LDA \textit{ab initio} values. The solid lines are LDA results, whereas the dashed lines are GGA counterparts.}
\label{missinglabel}
\end{minipage}
\\
\\
\begin{minipage}{\linewidth}
\centering
\includegraphics[scale=0.6]{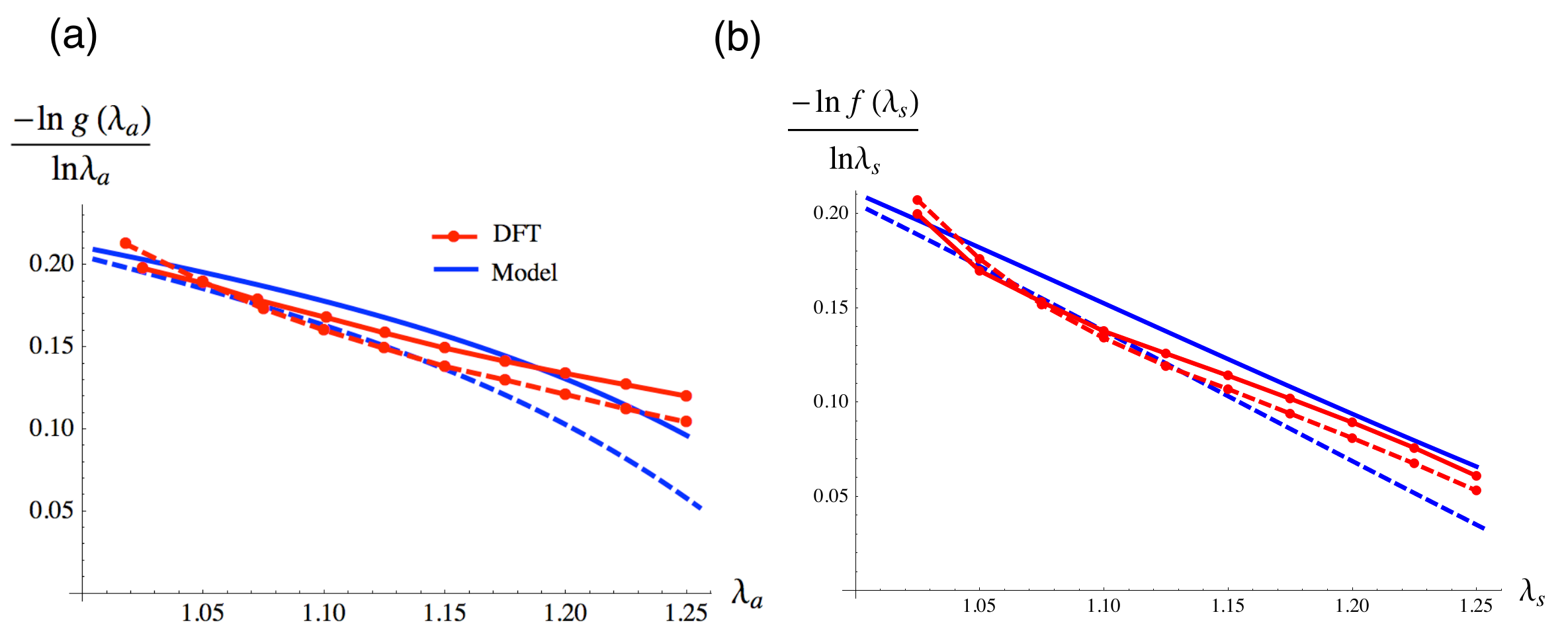}
\captionof{figure}{ Ratio of transverse strain (due to Poisson contraction) to longitudinal strain  for uniaxial stress along (a) the armchair direction and (b) the zigzag direction. The solid/dashed lines are LDA/GGA results. }
\label{poissonlike}
\end{minipage}
\section{Discussion and conclusion}
\label{summ}
Using as a basis the scalar-valued functions of the logarithmic strain tensor, called symmetry invariants, that remain invariant w.r.t. the point group symmetry of the graphene lattice, we derived a  nonlinear  hyperelastic constitutive response for graphene. Because the model employs symmetry invariants, the material symmetry group of the underlying lattice is built-into the model, and the need for externally imposing the symmetry restrictions is eliminated. This constitutive model is strictly hyperelastic, in the sense that the constants in the model are determined by fitting energies of the deformed states, without consideration of stresses. The model is coordinate-frame-independent,
 making it easy-to-implement in computational codes. The formulation allows straightforward evaluation of higher-order tensor variables such as the work-conjugate tangent moduli tensor and the acoustic tensor. The model clearly elucidates the contributions to the strain energy density due to purely equi-biaxial area change, and due to purely isochoric shape-changing deformations. The model correctly predicts the stress-strain variation of  graphene in cases of both uniaxial stretching and tension along both the zigzag and armchair directions, and in biaxial tension. The  values of the isotropic small strain elastic constants deduced from the model are also in good agreement with the measured values. The acoustic-tensor-based stability analysis predicts  failure stretches that are in good agreement with  independent lattice dynamics calculations based on linear response perturbation theory.

Our model predicts that the initial loss of elastic stability under  pure biaxial stretch occurs 
at $\epsilon_a \doteq 0.35$, by a vanishing tangent shear modulus, rather than being due to the vanishing tangent area modulus at $\epsilon_a \doteq 0.40 - 0.42$.  The latter instability mode was previously-reported by Wei \textit{et al.}, but we find that it could occur only at a larger deformation than that of the presently-identified elastic shear instability.
And as noted previously by Yevick  \& Marianetti \cite{marianetti2010failure},  phonon
 calculations also show that, prior to the onset of this long-wavelength shear instability, a short-wavelength instability at $K$ emerges earlier, at $\epsilon_a =  0.28 -0.30$, as seen also in Fig.~(\ref{acoustic}-c). Thus, the strength of graphene under  equi-biaxial tension is strictly limited by this short-wavelength instability. 

\end{document}